\documentclass[10pt,final,twocolumn]{IEEEtran}
\usepackage{ amsmath,graphicx}
\usepackage{tabularx,booktabs}
\newcolumntype{C}{>{\centering\arraybackslash}X}
\usepackage{lipsum}
\usepackage{algorithmicx}
\usepackage{algorithm}
\usepackage{algpseudocode}
\usepackage{amsfonts}
\usepackage{amsmath}
\usepackage{amssymb}
\usepackage{bigstrut}
\usepackage{bbm}
\usepackage{bm}
\usepackage{caption}
\usepackage{cite}
\usepackage{color}
\usepackage{colortbl}
\usepackage{eqparbox}
\usepackage{epstopdf}
\usepackage{epsfig}
\usepackage{epsf}
\usepackage{float}
\usepackage{footnote}
\usepackage{graphicx}
\usepackage{latexsym}
\usepackage{mwe}
\usepackage{mdwmath}
\usepackage{mdwtab}
\usepackage{placeins}
\usepackage{rotating}
\usepackage{stackengine}
\usepackage{subcaption}
\usepackage{syntonly}
\usepackage{tabularx}
\usepackage{times}
\usepackage{url}
\usepackage{verbatim}
\usepackage{wrapfig}
\usepackage{xspace}
\usepackage{epstopdf}
\usepackage{algpseudocode}
\usepackage{algorithmicx}
\usepackage{algorithm}

\def \ba {{\mathbf a}}

\def \bp {{\mathbf p}}
\def \bs {{\mathbf s}}

\begin{document}

\title{Optimal and Scalable Caching for 5G Using Reinforcement Learning of Space-time Popularities}

\author{
		Alireza~Sadeghi,
	Fatemeh Sheikholeslami, 
	and Georgios B. Giannakis\\ 

	\IEEEauthorblockA {Dept. of ECE and Digital Tech. Center, University of Minnesota}\\
{	Minneapolis, MN 55455, USA}\\
{	E-mails: \{sadeg012,sheik081,georgios\}@umn.edu}

	\thanks{	This work was supported by NSF 1423316, 1508993, 1514056, 1711471. 
}}

%

\maketitle

\begin{abstract}

	Small basestations (SBs) equipped with caching units have potential to handle the unprecedented demand growth in heterogeneous networks. Through low-rate, backhaul connections with the backbone, SBs can prefetch popular files during off-peak traffic hours, and service them to the edge at peak  periods. To intelligently prefetch, each SB must learn what and when to cache, while taking into account SB memory limitations, the massive number of available contents, the unknown  popularity profiles, as well as the space-time popularity dynamics of user file requests. In this work, local and global Markov processes model user requests, and a reinforcement learning (RL) framework is put forth for finding the optimal caching policy when the transition probabilities involved are unknown. Joint consideration of global and local popularity demands along with cache-refreshing costs allow for a simple, yet practical asynchronous caching approach. The novel RL-based caching relies on a Q-learning algorithm to implement the optimal policy in an online fashion, thus enabling   the cache control unit at the SB to learn, track, and possibly adapt to the underlying dynamics. 
	To endow the algorithm with scalability, a linear function approximation of the proposed Q-learning scheme is introduced, offering faster convergence as well as reduced complexity and memory requirements. Numerical tests corroborate the merits of the proposed approach in various realistic settings.

\end{abstract}

\begin{IEEEkeywords}
	Caching, dynamic popularity profile, reinforcement learning, Markov decision process (MDP), Q-learning.
\end{IEEEkeywords}

\IEEEpeerreviewmaketitle

\section{Introduction}

The advent of smart phones, tablets, mobile routers, and a massive number of devices connected through the Internet of Things (IoT) have led to an unprecedented growth in data traffic. Increased number of users trending towards video streams, web browsing, social networking and online gaming, have urged  providers to pursue new service technologies that offer acceptable quality of experience (QoE). One such technology entails network densification by deploying small pico- and femto-cells, each serviced by a low-power, low-coverage, small basestation (SB). In this infrastructure,  referred to as  heterogeneous network (HetNet), SBs are connected to the backbone by a cheap `backhaul' link.  While boosting the network density by substantial reuse of scarce resources, e.g., frequency, the HetNet architecture is restrained by its low-rate, unreliable, and relatively slow backhaul links~\cite{Femtocell_Andrews_2012}.   

During peak traffic periods specially when electricity prices are also high, weak backhaul links can easily become congested--an effect lowering the QoE for end users. One approach  to mitigate this limitation is to shift the excess load from peak periods to off-peak periods. Caching realizes this shift by fetching the ``anticipated'' popular contents, e.g., reusable video streams, during off-peak periods, storing this data in SBs equipped with memory units, and reusing them during peak traffic hours~\cite{Wireless_caching_Paschos_2016,Femrocaching_Golrezaei_2013, Cacheinair}. 
In order to utilize the caching capacity intelligently, a content-agnostic SB must rely on available observations to learn what and when to cache. To this end, machine learning tools can provide 5G cellular networks with efficient caching, in which a ``smart'' caching control unit (CCU) can learn, track, and possibly adapt to the space-time popularities of reusable contents~\cite{Wireless_caching_Paschos_2016,What5Gbe_Andrews_2014}. 

\noindent{\bf Prior work}. Existing efforts in 5G caching  have focused on enabling SBs to learn unknown time-invariant  content popularity profiles, and cache the most popular ones accordingly. A multi-armed bandit approach is reported in \cite{MAB_Gunduz_2014}, where a reward is received when user requests are served via cache; see also \cite{LearnDist_Sengupta_2014} for a distributed, coded, and convexified reformulation. 
%
%
%
A belief propagation-based approach for distributed and collaborative caching is also investigated in \cite{DistBeliefprop}. Beyond \cite{MAB_Gunduz_2014}, \cite{LearnDist_Sengupta_2014}, and \cite{DistBeliefprop} that deal with deterministic caching, \cite{JointOptimal} and \cite{OptimalGeographic} introduce probabilistic alternatives. Caching, routing and video encoding are jointly pursued in \cite{VideoDeliveryOpt} with users having different QoE requirements. However, a limiting assumption in \cite{MAB_Gunduz_2014,LearnDist_Sengupta_2014,DistBeliefprop,JointOptimal,
	OptimalGeographic,VideoDeliveryOpt} pertains to space-time invariant modeling of  popularities, which can only serve as a crude approximation for real-world requests. Indeed, temporal dynamics of local requests are prevalent due to  user mobility, as well as emergence of new contents, or, aging of older ones.
To accommodate dynamics,   Ornstein-Uhlenbeck processes and Poisson shot noise models are utilized in  \cite{Dynamic_Debbah_2017} and \cite{leconte2016}, respectively, while context- and trend-aware caching approaches are investigated in \cite{Context_aware_Schaar_2017} and \cite{Trend_aware_Schaar_2016}.

Another practical consideration for 5G caching is driven by the fact that a relatively small number of users request contents during a caching period. This along with the small size of cells can challenge SBs from estimating accurately the underlying content popularities.
To address this issue, a  transfer-learning approach  is advocated in  \cite{Transferlearning_Debbah_2015}, \cite{Learning_Poor_2016} and  \cite{leconte2016}, to improve the time-invariant popularity profile estimates by leveraging prior information obtained from a surrogate (source) domain, such as social networks.


Finally, recent studies have investigated the role of coding for enhancing performance in cache-enabled networks \cite{Maddahalifundamentals,Maddahalidecent,MaddahaliOnline}; see also \cite{CaireFundamentalsD2D}, \cite{CaireWD2Dcaching}, and \cite{DtoD}, where device-to-device ``structureless'' caching approaches are envisioned.

\noindent{\bf Contributions}. The present paper introduces a novel approach to account for space-time popularity of user requests by casting the caching task in a reinforcement learning (RL) framework. The CCU of the local SB is equipped with storage and processing units for solving  the emerging RL optimization in an online fashion. Adopting a Markov model for the popularity dynamics, a Q-learning caching algorithm is developed to learn the optimal policy when the underlying transition probabilities are unknown.   

Given the geographical and temporal variability of cellular traffic, global popularity profiles may not always be representative of local demands. To capture this, the proposed framework entails estimation of the popularity profiles  both at the local as well as at the global scale. Specifically, each SB estimates its local vector of popularity profiles based on limited observations, and transmits it to the network operator, where an estimate of the global profile is obtained by  aggregating the local ones. The estimate of the global popularity vector is then  sent back to the SBs. The SBs can adjust the cost (reward) to  trade-off tracking global trends versus serving local requests.

To obtain a scalable caching scheme, a novel approximation of the proposed Q-learning algorithm is also developed. Furthermore, despite the stationarity assumption on the popularity Markov models, proper selection of stepsizes broadens the scope of the proposed algorithms for  tracking  demands even in non-stationary settings. 

The rest of this paper is organized as follows. Section \ref{sec:model} introduces the system model and problem formulation, and Section \ref{Bellman} presents the optimal Q-learning algorithm. Linear function approximation, and the resultant scalable solver are the subjects of Section IV, while Section V presents numerical tests, and Section VI concludes the paper.

\emph{Notation.} Lower- (upper-) case boldface letters denote column vectors (matrices), whose $(i,j)$-th entry is denoted by $[\,.\,]_{i,j}$. Calligraphic symbols are reserved for sets, while $\top$ stands for transposition.

\section{Modeling and problem statement}\label{sec:model}
Consider a local section of a HetNet with  a single SB connected to the backbone network through a low-bandwidth, high-delay backhaul link. Suppose further that the SB is equipped with $M$ units to store contents (files) that are assumed for simplicity to have unit size; see Fig.~1. Caching will be carried out in a slotted fashion over slots $t=1,2,\ldots$,{ where at the end of each slot, the CCU-enabled SB selects ``intelligently'' $M$ files from the total of $F \gg M$ available ones at the backbone, and prefetches them for possible use in subsequent slots. 
	
The structure of every time slot is depicted in Fig. \ref{slot}. Specifically, at the beginning of every time slot, the user file requests are revealed and the ``content delivery" phase takes place. The second phase pertains to ``information exchange,'' where the SBs transmit their locally-observed popularity
	profiles to the network operator, and in return receive the estimated global popularity profile. Finally,
	``cache placement'' is carried out and optimal selection of files are stored for the next time slot.} The slots may not be of  equal length, as the starting times  may be set a priori, for example at 3~AM, 11~AM, or 4~PM, when the network load is low; or, slot intervals may be dictated to CCU by the network operator on the fly. Generally, a slot starts when the network is at an off-peak period, and its duration coincides with the peak traffic time when there is  pertinent costs of serving users are high.

During content delivery phase of slot $t$, each user locally requests a subset of  files from the set ${\cal F}:=\left\{1,2,\ldots,F\right\}$. If a requested file has been stored in the cache, it will be simply served locally, thus  incurring (almost) zero cost.  Conversely, if the requested file is not available in the cache, the SB must fetch it from  the cloud  through its cheap backhaul link, thus incurring a  considerable cost due to possible electricity price surges, processing cost, or the sizable delay resulting  in low QoE and user dissatisfaction.  The CCU wishes to intelligently select the cache contents so that costly services from the cloud be avoided as often as possible.

Let ${\bf a} (t) \in \mathcal{A}$ denote the $F \times 1$ binary \emph{caching action vector} at slot $t$, where $\mathcal{A}:=\{\mathbf{a}| \mathbf{a} \in \{0,  1\}^F, \mathbf{a}^\top \mathbf{1}=M\}$ is the set of all feasible actions; that is,  $[{{\bf a}(t)}]_f  = 1$ indicates that file $f$ is cached for the duration of slot $t$, and  $[{\bf a}(t)]_f  = 0$ otherwise.

\begin{figure}[t]\label{fig:sys1}
	\centering
	\includegraphics[width=0.9\columnwidth]{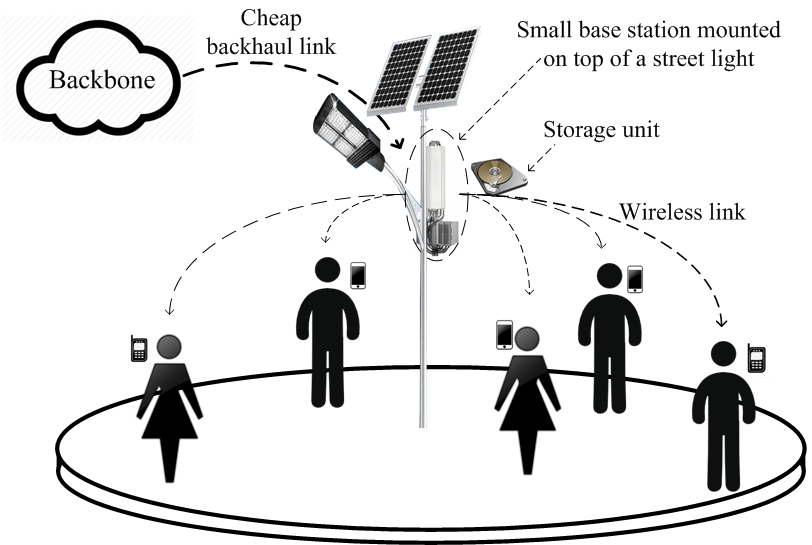}
	\vspace{0.2 cm}
	\caption{Local section of a HetNet.}
\end{figure}

Depending on the received requests from locally connected users during content delivery phase, the CCU computes the $F \times 1$-vector of \emph{local popularity profile} ${\bf p}_L (t)$ per slot $t$, whose $f$-th entry indicates  the expected local demand for file $f$, defined  as 
\begin{align}
\nonumber
\bigg[\mathbf{p}_{\text{L}} (t)\bigg]_f :=\dfrac{\text{ Number of local requests for } f \; {\text {at slot} } \;t }{\text{Number of all local requests at slot}\; t}.
\end{align}
Similarly, suppose that the backbone network estimates  the $F \times 1$ \emph{global popularity profile} vector ${\bf p}_G (t)$, and transmits it to~all~CCUs. 

Having observed the local and global user requests by the end of information exchange phase of slot $t$, our overall system state is 
\begin{equation}
{\bf s} (t):= \left[{\bf p}^{\top}_G (t),{\bf p}^{\top}_L (t),{\bf a}^{\top}(t)\right]^{\top}.
\label{eq.1}
\end{equation}
Being at slot $t-1$, our \emph{objective} is to leverage historical observations of states, $\left\{{\bf s} (\tau)\right\}_{\tau = 0}^{t-1}$, and pertinent costs in order to learn the optimal action for the next slot, namely ${\bf a}^{\ast}(t)$. Explicit expression of the incurred costs, and analytical formulation of the objective will be elaborated in the ensuing subsections.

\subsection{Cost functions and caching strategies}
\label{Regorously_Problem_formulation}
{Efficiency of a caching strategy will be measured by how well it utilizes the available storage of the local SB to keep the most popular files, versus how often local user requests are met via fetching through the more expensive backhaul link. The overall cost incurred will be modeled as the superposition of three types of costs.

\begin{subequations}	
The first type $c_{1,t}$ corresponds to the cost of refreshing the cache contents. In its  general form, $c_{1,t} (\cdot)$ is a function of the upcoming action $\mathbf{a}(t)$, and available contents at the cache according to current caching action $\mathbf{a}(t-1)$, where the subscript $t$ captures the possibility of a time-varying cost for refreshing the cache. A reasonable choice of $c_{1,t}(\cdot)$ is
\begin{equation}
c_{1,t}(\mathbf{a}(t),\mathbf{a}(t-1)) := \lambda_{1,t} \mathbf{a}^\top(t) \left[\mathbf{1}- \mathbf{a}(t-1)\right] 
\label{subcost1}
\end{equation}
which upon recalling that the action vectors $\mathbf{a}(t-1)$ and $\mathbf{a}(t)$ have binary $\left\{0,1\right\}$ entries, implies that $c_{1,t}$ counts the number of those files to be fetched and cached  prior to slot $t$, which were not stored according to action $\mathbf{a}(t-1)$.

The second type of cost is incurred during the operational phase of slot $t$ to satisfy user requests.
With $c_{2,t}(\mathbf{s}(t))$ denoting this type of cost, a prudent choice must: i)~penalize requests for files already cached much less than requests for files not stored; and, ii) be a non-decreasing function of popularities $[\mathbf{p}_L]_f$. Here for simplicity, we assume that the transmission cost of cached files is relatively negligible, and choose 
\begin{equation}
c_{2,t}(\mathbf{s}(t)):= \lambda_{2,t} \left[\mathbf{1}-\mathbf{a}(t)\right]^\top \mathbf{p}_L(t)
\label{subcost2}
\end{equation}
which solely penalizes the non-cached files in descending order of their local {popularities}.
	
The third type of  cost captures the  ``mismatch'' between caching action $\mathbf{a}(t)$, and the global popularity profile $\mathbf{p}_{\text{G}}(t)$. Indeed, it is  reasonable to consider the global popularity of files as an acceptable representative of what the local profiles will look like in the near future; thus, keeping the caching action close to $\mathbf{p}_{\text{G}}(t)$ may reduce future possible costs. Note also that a relatively small number of local requests may only provide a crude estimate of local popularities, while the global popularity profile can serve as side information in tracking the evolution of content popularities over the network. { Moreover, in the networks with highly mobile users storing globally popular files, might be a better caching decision than the local ones.} This has prompted the advocation of transfer learning approaches, where content popularities in a surrogate domain are utilized for improving estimates of popularity; see, e.g., \cite{Transferlearning_Debbah_2015} and \cite{Learning_Poor_2016}. However, this approach is limited by the degree the surrogate (source) domain, e.g., Facebook or Twitter, is a good representative of the target domain requests. When it is not, techniques will  misguide caching decisions, while imposing excess  processing overhead to the network operator or to the SB. 

To account for this issue, we introduce the third type of cost as
\begin{equation}
c_{3,t}(\mathbf{s}(t)):= \lambda_{3,t} \left[\mathbf{1}-\mathbf{a}(t)\right]^\top \mathbf{p}_G(t)
\label{subcost3}
\end{equation}
 penalizing the files not cached according to the global popularity profile  ${\bf p}_G (\cdot)$ provided by the  network operator, thus promoting adaptation of caching policies close to global demand~trends.

\end{subequations}

All in all, upon taking action $\mathbf{a}(t)$ for slot $t$, the \emph{aggregate cost conditioned} on the popularity vectors revealed, can be expressed as (cf. \eqref{subcost1}-\eqref{subcost3}) 
\begin{align}\label{Overall_Cost}
& C_t \Big({ {\bf s} (t-1), {\bf a} (t) \Big| {\mathbf{p}_{\text{G}}(t)},{\mathbf{p}_{\text{L}}(t)}} \Big) \\ & \nonumber  \hspace{1cm} :=   c_{1,t}\left({\bf a} (t), {\bf a} (t-1)\right) + c_{2,t}\left({\bf s} (t)\right) + c_{3,t}(\mathbf{s}(t)) \\ \nonumber &  \hspace{1.1cm}= \lambda_{1,t} \mathbf{a}^\top(t) (\mathbf{1}- \mathbf{a}(t-1)) +\lambda_{2,t} (\mathbf{1}-\mathbf{a}(t))^\top \mathbf{p}_L(t)  \\& \hspace{1.1cm}+ \lambda_{3,t} (\mathbf{1}-\mathbf{a}(t))^\top \mathbf{p}_G(t).\nonumber
\end{align} 
Weights $\lambda_{1,t}$, $\lambda_{2,t}$, and $\lambda_{3,t}$ control the relative significance of  the corresponding summands, whose tuning influences the optimal caching policy at the CCU. As asserted earlier, the cache-refreshing cost at off-peak periods is considered to be less than fetching the contents during  slots, which justifies the choice $\lambda_{1,t} \ll \lambda_{2,t}$. In addition, setting $\lambda_{3,t} \ll \lambda_{2,t}$ is of interest when  the local popularity profiles are of acceptable accuracy, or, if tracking local popularities is of higher importance. In particular, setting $\lambda_{3,t} = 0$ corresponds to the special case where the caching cost is  decoupled from the global popularity profile evolution. On the other hand, setting $\lambda_{2,t} \ll \lambda_{3,t}$  is desirable in networks where globally popular files are of high importance, for instance when
users have high mobility and may change SBs rapidly, or, when a few local requests prevent the SB from estimating accurately the local popularity profiles. Fig.~\ref{fig:sys2} depicts the evolution of popularity and action vectors along with the aggregate conditional costs across slots.

\begin{figure}[t] 
	\centering
	{\includegraphics[width=1\columnwidth]{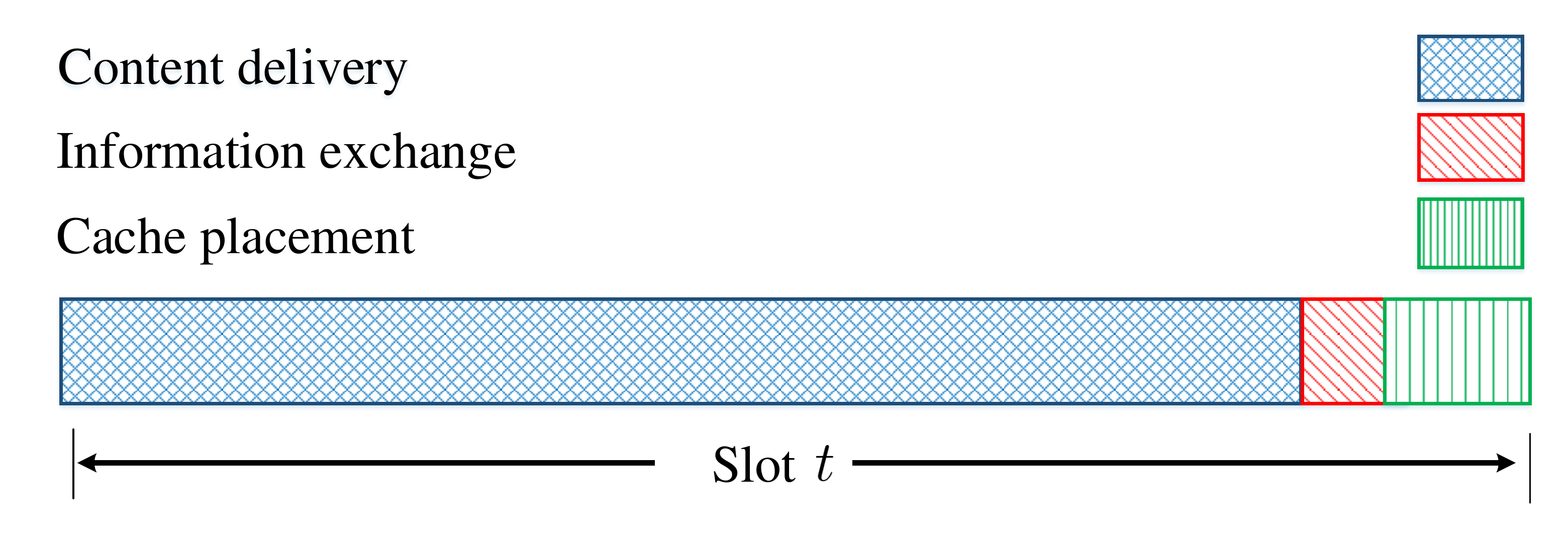}}
	\caption{The slot structure.} 
	\label{slot} 	
\end{figure}

\begin{figure}[t] 
	\centering
	{\includegraphics[width=1\columnwidth]{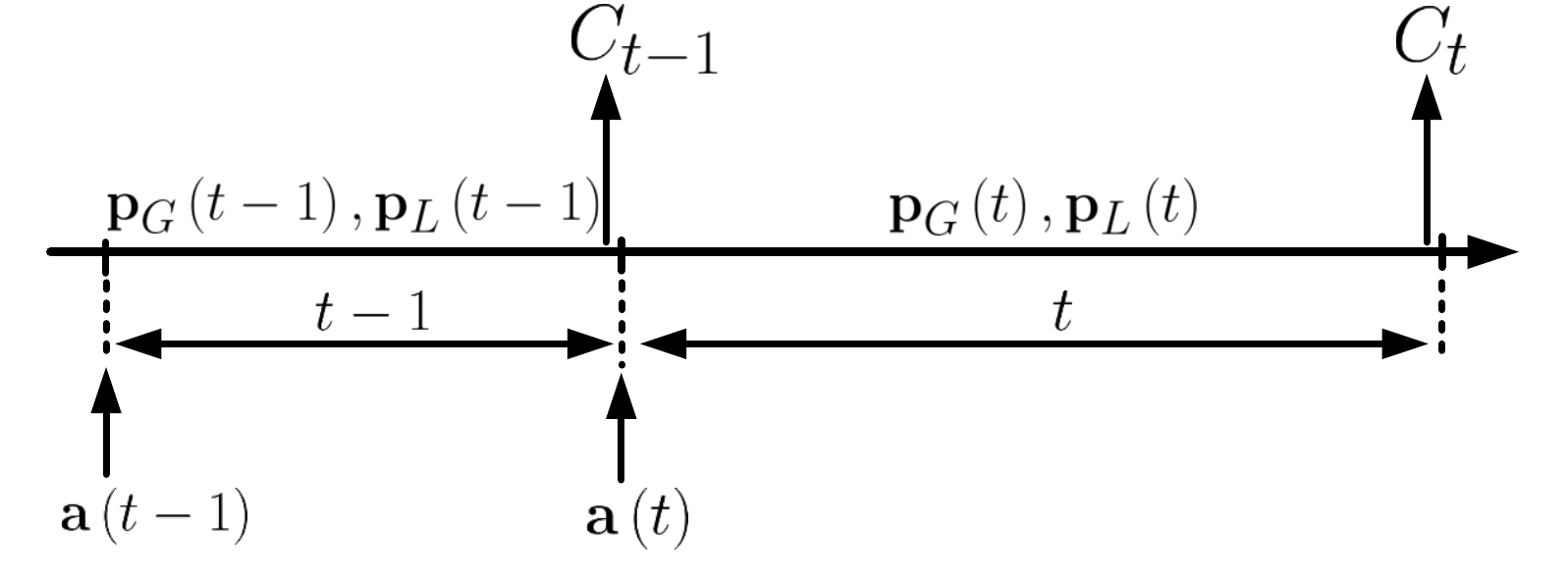}}
	\caption{A schematic depicting the evolution of key quantities across time slots. Duration of slots can be unequal.} 
	\label{fig:sys2} 	
\end{figure}

\noindent{\bf Remark 1.} As with slot sizes, proper selection of $ \lambda_{1,t}, \lambda_{2,t}$, and $\lambda_{3,t}$ is a design choice, {and is assumed to be dictated by the   network operator at information exchange phase within each time  slot. To provide these parameters, the network operator must observe user request patterns, the electricity price, and global and local popularity mismatch of different SBs during a typical day, and then provide the CCUs with cost parameters $\lambda_1$, $\lambda_2$, and $\lambda_3$, accordingly. For instance, cache refreshing cost $\lambda_1$ can be set low during night as electricity price is very low, and during hours of the day when users are often mobile, e.g.,
commuting, the global popularity mismatch parameter $\lambda_3$ should be higher than the local one $\lambda_2$.}

Indeed, the overall approach requires the network service provider and the SBs  to inter-operate by exchanging relevant information. Estimation of global popularities  requires SBs to transmit their locally obtained ${\bf p}_L(t)$  to the network operator at information exchange phase of each slot. { Subsequently}, the network operator informs the CCUs of the (estimated) global popularity ${\bf p}_G(t)$, and cost parameters $\lambda_{1,t}, \lambda_{2,t}$, and $\lambda_{3,t}$.  By providing the network operator with means of parameter selection, a
``master-slave'' hierarchy emerges, which enables the network operator (master)  to influence SBs (slaves) caching decisions, leading to a  centrally controlled adjustment of caching policies. Interestingly, these  few bytes of information exchanges occur once per slot and at  off-peak instances, thus imposing  negligible overhead to the system, while enabling a simple, yet practical and powerful optimal semi-distributed caching process; see Fig.~\ref{fig:sys3}.


\subsection{Popularity profile dynamics}
\label{Dynamic evolution of user demands}
As depicted in Fig.~3, we will model user requests (and thus popularities) at both global and local scales using Markov chains. Specifically,  global popularity profiles will be assumed generated by an underlying Markov process with $|{\cal P}_G|$ states collected in the set  \linebreak
${\cal P}_G :=\left\{ {\bf p}_{G}^{1}, \ldots, {\bf p}_{G}^{|{\cal P}_G|} \right\}$ ; and likewise for the set of all local popularity profiles 
${\cal P}_L:=\left\{ {\bf p}_{L}^{1}, \ldots, {\bf p}_{L}^{|{\cal P}_L|} \right\}$. Although ${\cal P}_G$ and ${\cal P}_L$ are known, the underlying transition probabilities of the two Markov processes are considered unknown. 

Given ${\cal P}_G$ and ${\cal P}_L$ as well as feasible caching decisions in set $\cal A$, the overall set of states in the network is 
\begin{align}
\nonumber {\cal S} := \left\{{\bf s}| {\bf s}= [{\bf p}^{\top}_G, {\bf p}^{\top}_L, {\bf a}^{\top}]^{\top}, {\bf p}_G \in {\cal P}_G  \textrm{ , } {\bf p}_L \in {\cal P}_L, {\bf a} \in {\cal A} \right\}.
\end{align} 
{In the proposed RL based caching, the underlying transition probabilities for global and local popularity profiles are considered  \textit{unknown}, which is a practical assumption. } In this approach, the learner seeks the optimal policy by interactively making sequential decisions, and observing the corresponding costs. The caching task is formulated in the following subsection,  and an efficient solver is developed to cope with the ``curse of dimensionality'' typically emerging with RL problems \cite{Sutton1998reinforcement}.

\begin{figure}[t]
	\centering
	{\includegraphics[width=1.12\columnwidth]{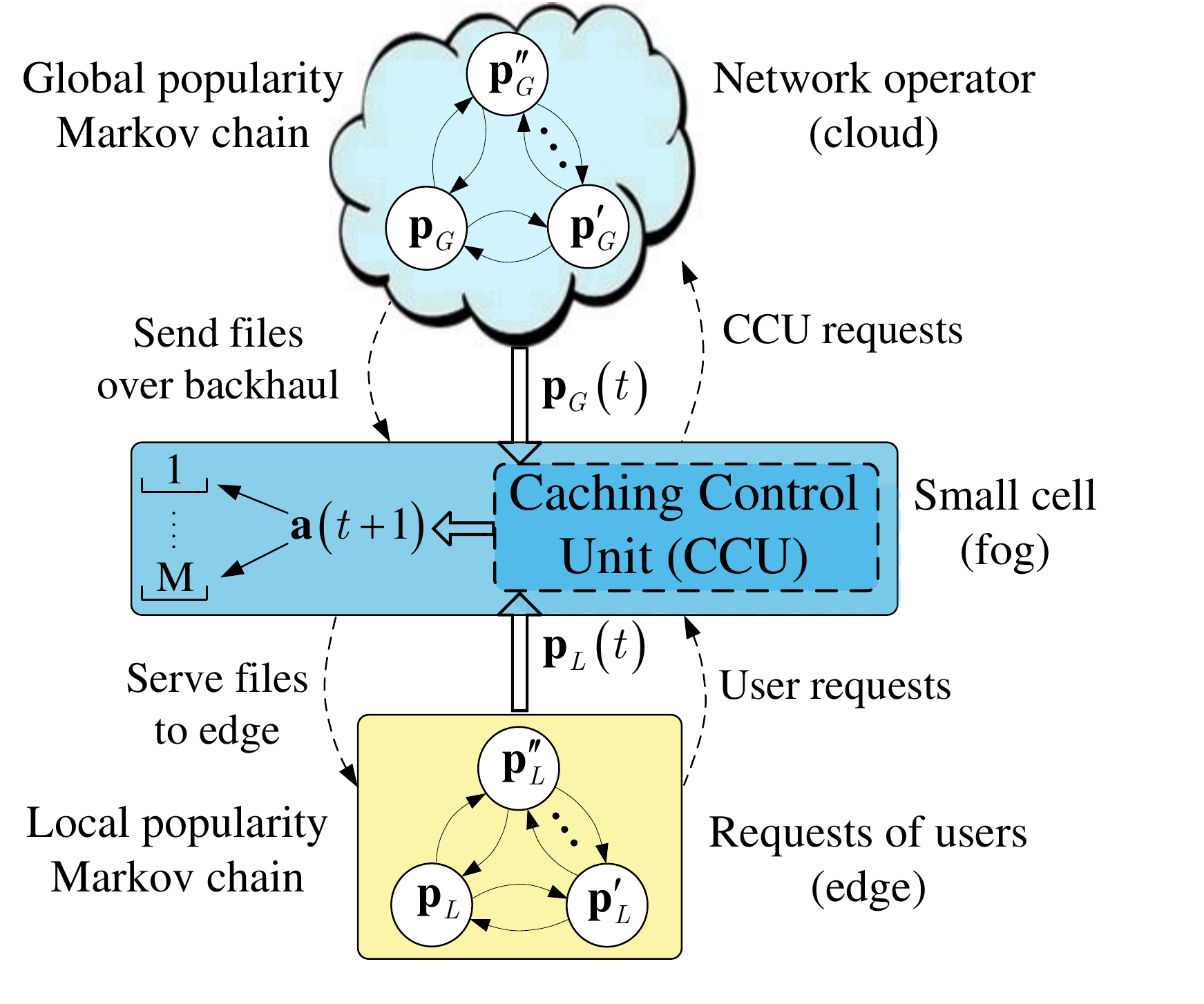}}
	\caption{Schematic of network structure with required communications in SBs, and with the network operator.} 
	\label{fig:sys3} 
\end{figure}
\subsection{Reinforcement learning formulation}	

As showing in Fig.~3, at the end of time slot $t-1$ the CCU takes caching action $\mathbf{a}(t)$ to meet the upcoming requests, and by the end of content delivery as well as information exchange  phases of  slot $t$, the profiles $\mathbf{p}_G(t)$ and $\mathbf{p}_L(t)$ become available, so that the system state is updated to $\mathbf{s}(t)$, and the conditional cost $C_t \Big( { {\bf s} (t-1), {\bf a} (t) {\Big |} {\mathbf{p}_{\text{G}}(t)},{\mathbf{p}_{\text{L}}(t)}} \Big)$ is revealed.	
Given the random nature of user requests locally and globally, $C_t$ in \eqref{Overall_Cost} is a random variable with mean	
\begin{align}\label{mean_Cost}
& {\overline C}_t \left( {\bf s} (t-1), {\bf a} (t) \right)   \\
& :=\mathbb{E}_{{\mathbf{p}_{\text{G}}(t)},{\mathbf{p}_{\text{L}}(t)}} \Big[ C_t \left( { {\bf s} (t-1), {\bf a} (t) {\Big |} {\mathbf{p}_{\text{G}}(t)},{\mathbf{p}_{\text{L}}(t)}} \right) \Big] \nonumber \\
&  = \lambda_1 \mathbf{a}^\top(t) \left[\mathbf{1}- \mathbf{a}(t-1)\right] +\lambda_2 \mathbb{E} \left[ (\mathbf{1}-\mathbf{a}(t))^\top \mathbf{p}_L(t)\right] \nonumber \\  & + \lambda_3 \mathbb{E} \left[(\mathbf{1}-\mathbf{a}(t))^\top \mathbf{p}_G(t)\right] \nonumber
\end{align}
where the expectation is taken with respect to (wrt)  ${\bf p}_L (t)$ and ${\bf p}_G (t)$, while the weights are selected as $\lambda_{1,t} = \lambda_1$, $\lambda_{2,t} = \lambda_2$, and $\lambda_{3,t} = \lambda_3$ for simplicity.

Let us now define the policy function $\pi: \mathcal{S} \rightarrow \mathcal{A}$, which maps any state $\mathbf{s} \in  \mathcal{S}$ to the action set. Under policy $\pi(\cdot)$, for the current state ${\bf s}(t)$,  caching is carried out via action $\mathbf{a}(t+1)=\pi(\mathbf{s}(t))$  dictating what files to be stored for the $(t+1)$-st slot. Caching performance is measured through the so-termed state value function 
\begin{align}\label{eq_Value_function}
 V_{\pi} \left({\bf s} (t)\right) :=   \lim\limits_{T \rightarrow \infty} \mathbb{E}  \left[\sum\limits_{\tau = t}^{T} \gamma^{\tau-t} {\overline C}\left({\bf s} \left[\tau\right], {\pi} \left(\mathbf{s}\left[\tau\right]\right) \right)\right] 
\end{align}
 which is the total average cost incurred over an infinite time horizon, with future terms discounted by factor  $\gamma \in \left[0,1\right)$. Since taking action ${\bf a}(t)$ influences the SB state in future slots, future costs are always affected by past and present actions. Discount factor $\gamma$ captures this effect, whose tuning trades  off current versus future costs. Moreover, $\gamma$ also accounts for modeling uncertainties, as well as imperfections, or dynamics. For instance, if there is ambiguity about future costs, or if the system changes very fast, setting  $\gamma$ to a small value enables one to prioritize current costs, whereas in a stationary setting one may prefer to demote future costs through a larger $\gamma$. 

The objective of this paper is to find the optimal policy $\pi^*$ such that  the average cost of any state ${\bf s}$ is minimized  (cf. \eqref{eq_Value_function}) 
\begin{equation}
\label{eq_Opt_policy}
{\pi^{\ast}} = \arg \min \limits_{\pi \in \Pi} V_{\pi} \left( {\bf s} \right), \quad \forall {\bf s} \in {\cal S}
\end{equation} where $\Pi$ denotes the set of all feasible policies.

 The  optimization in  \eqref{eq_Opt_policy} is a sequential decision making problem. In the ensuing section, we present optimality conditions (known as Bellman equations) for our problem, and  introduce a~{Q-learning} approach for solving \eqref{eq_Opt_policy}.}

\section{Optimality conditions}
\label{Bellman}
Bellman equations, also known as dynamic programming equations, provide necessary conditions for optimality of a policy in a sequential decision making problem. Being at the $(t-1)$st slot, let  $[{\bm P}^a]_{{\bf s}{\bf s}'}$ denote the transition probability of going from the current state ${\bf s}$ to the next state ${\bf s}'$ under action ${\bf a}$; that is, \[ [{\bm P}^a]_{{\bf s}{\bf s}'} := {\rm{Pr}} \Big \{{{\bf s}(t) = {\bf s}'}  {\Big |} {\bf s}(t-1) = {\bf s}, \pi(\mathbf{s}(t-1)) = \mathbf{a}\Big \}. \]
Bellman equations express the state value function by \eqref{eq_Value_function} in a recursive fashion as \cite[pg. 47]{Sutton1998reinforcement}
\begin{equation}\label{eq_Value_function_recurring}
V_{\pi} \left({\bf s} \right) = { \overline{C}\left({\bf s}, \pi(\mathbf{s})  \right)  }+ \gamma  \sum_{{\bf s}' \in \mathcal{S}} [{\bm P}^{\pi({\bf s})}]_{{\bf s}{\bf s}'}   V_{\pi} \left({\bf s}' \right)\;, \forall {\bf s,s}'
\end{equation}
which amounts to the superposition of $\overline C$ plus a discounted version of future state value functions under a given policy $\pi$. Specifically, after dropping the current slot index $t-1$ and indicating with prime quantities of the next slot $t$,  ${ \overline{C}}$ in \eqref{mean_Cost} can be written as
\begin{align*}
\label{cbardef}
{ \overline{C}\left({\bf s}, \pi(\mathbf{s})  \right)  } = \sum_{\bs'  := [\bp_G',\bp_L',\ba']\in \mathcal{S}}  [{\bm P}^{\pi({\bf s})}]_{{\bf s}{\bf s}'}  C\Big({\bf s}, \pi(\mathbf{s}) \Big| \bp_G',  \bp_L'   \Big)  
\end{align*}
where $ C\Big({\bf s}, \pi(\mathbf{s}) \Big| \bp_G',  \bp_L'   \Big)$ is found as in \eqref{Overall_Cost}.
It turns out that, with $[{\bm P}^a]_{{\bf s}{\bf s}'}$ given $\forall {\bf s,s}'$, one can readily obtain $\left\{V_{\pi} ({\bs}), \forall {\bs} \right\}$ by solving \eqref{eq_Value_function_recurring}, and eventually the optimal policy $\pi^{\ast}$ in \eqref{pi*} using the so-termed policy iteration algorithm \cite[pg. 79]{Sutton1998reinforcement}. To outline how this algorithm works in our context, define the state-action value function that we will rely on under policy $\pi$ \cite[pg. 62]{Sutton1998reinforcement}
\begin{equation}
\label{Q_pi}
Q_{\pi} \left({\bf s} , {{\ba}'}\right) :=    { \overline{C}\left({\bf s},  {\ba'}  \right)}  +  \gamma  \sum_{{\bf s}' \in \mathcal{S}} [{\bm P}^{{\ba'}}]_{{\bf s}{\bf s}'}   V_{\pi} \left({\bf s}' \right). 
\end{equation}
Commonly referred to as the ``Q-function,'' $Q_{\pi}({\bf s},{\boldsymbol \alpha})$ basically captures the expected current cost of taking action ${\boldsymbol \alpha}$ when the system is in state $\bf s$, followed by the discounted value of the future states, provided that the future actions are taken according to policy~$\pi$.

 In our setting, the policy iteration algorithm initialized with $\pi_0$, proceeds with the following updates at the  $i$th iteration.
\begin{itemize}
	\item \textbf{Policy evaluation}: Determine $V_{\pi_i}(\bf s)$ for all states ${\bf s} \in {\cal S}$ under the current (fixed) policy $\pi_i$, by solving the system of linear equations in \eqref{eq_Value_function_recurring} $\forall {\bf s}$.
	\item \textbf{Policy update}: Update the policy using
\[ \pi_{i+1} ({\bf s}) := \arg \min_{{\boldsymbol \alpha}} Q_{\pi_{i}} ({\bf s},{\boldsymbol \alpha}), \quad \forall {\bf s} \in {\cal S}. \]
\end{itemize}

The policy evaluation step is of complexity~$\mathcal{O}(|\mathcal{S}|^3)$, since it requires  matrix inversion for solving  the linear system of equations in \eqref{eq_Value_function_recurring}. Furthermore, given $V_{\pi_i}(\bs) \; \forall \bs$, the complexity of the policy update step is  $\mathcal{O}(|{\cal A}||\mathcal{S}|^2)$, since the Q-values must be updated per state-action pair, each subject to~$|\cal S|$ operations;  see also \eqref{Q_pi}.
Thus, the per iteration complexity of the policy iteration algorithm is $\mathcal{O}(|\mathcal{S}|^3+|{\cal A}||\mathcal{S}|^2)$. Iterations proceed until convergence, i.e., $\pi_{i+1}({\bf s})={\pi_i({\bf s})}, \, \forall {\bf s}\in {\cal S}$.

Clearly, the policy iteration algorithm relies on knowing $[{\bm P}^{\bf a}]_{{\bf s}{\bf s}'}$, which is typically not available in practice. This motivates the use of adaptive dynamic programming (ADP) that learn $[{\bm P}^{\bf a}]_{{\bf s}{\bf s}'}$ for all ${\bf s, s}' \in \mathcal{S}$, and ${\bf a} \in \mathcal{A}$, as iterations proceed \cite[pg. 834]{AIModernapproach}. Unfortunately, ADP algorithms are often very slow and impractical, as they must estimate $|\mathcal{S} |^2 \times |\mathcal{A}|$ probabilities. In contrast, the Q-learning algorithm elaborated next finds the optimal $\pi^{*}$ as well as $V_{\pi}({\bf s})$, while circumventing the need to estimate $[{\bm P}^a]_{{\bf s}{\bf s}'}, \forall {\bf s,s}'$; see e.g., \cite[pg.~140]{Sutton1998reinforcement}. 

 \subsection{Optimal caching via Q-learning}
 \label{Q-learning}
 Q-learning is an online RL scheme to jointly infer the optimal policy $\pi^{\ast}$, and estimate the optimal state-action value function  $Q^*(\mathbf{s,a}')  := Q_{\pi^{\ast}}(\mathbf{s,a}') \quad \forall {\bf s},{\bf a}'$. Utilizing \eqref{eq_Value_function_recurring} for the optimal policy $\pi^{\ast}$, it can be shown that \cite[pg. 67]{Sutton1998reinforcement} \linebreak
 \begin{equation}\label{pi*}
 \pi^*({\bf s}) = \arg\min_{{\boldsymbol \alpha}}~Q^{*}({\bf s}, {\boldsymbol \alpha}), \quad \forall {\bf s} \in {\cal S}.
 \end{equation}
 The Q-function and  $V(\cdot)$ under $\pi^{\ast}$ are related by
 \begin{equation}\label{v*}
 V^{*}({\bf s}):= V_{\pi^*}({\bf s}) =\min_{{\boldsymbol \alpha}} Q^{*}({\bf s}, {\boldsymbol \alpha})
 \end{equation}
 which in turn yields 
\begin{equation}
\label{eq_Q_function2}
 Q^* \left({\bf s} , {\bf a}'\right) =  \overline{C}\left({\bf s}, {\bf a}'  \right) +  \gamma  \sum_{{\bf s}' \in \mathcal{S}} [{\bm P}^{\bf a}]_{{\bf s}{\bf s}'}  \min_{{\boldsymbol \alpha}\in {\cal A}} Q^* \left({\bf s'} , {\boldsymbol \alpha}\right).
\end{equation} 

Capitalizing on the optimality conditions \eqref{pi*}-\eqref{eq_Q_function2}, an online Q-learning scheme for caching is listed under  Alg.~1. 
 In this algorithm, the agent updates its estimated $\hat{Q}(\mathbf{s}(t-1),\mathbf{a}(t))$ as  $C\Big( {{\bf s} (t-1), {\bf a} (t)} \Big |  {\mathbf{p}_{\text{G}}(t)},{\mathbf{p}_{\text{L}}(t)} \Big)$ is observed. That is, given ${\bf s}(t-1)$, Q-learning takes action ${\mathbf{a}}(t)$, and  upon observing ${\mathbf{s}}(t)$, it incurs cost $C\Big( {{\bf s} (t-1), {\bf a} (t)} \Big|  {\mathbf{p}_{\text{G}}(t)},{\mathbf{p}_{\text{L}}(t)} \Big)$.  Based on the instantaneous error 
\begin{align}\label{error}
\varepsilon \left({\bf s}(t-1), {\bf a}(t)\right) :=  & \frac{1}{2} \Big( C\left({\bf s}(t-1),{\bf a}(t)\right) + \gamma \min \limits_{{\boldsymbol \alpha}}^{} {\widehat Q} \left({\bf s}(t), {\boldsymbol \alpha}\right)  \nonumber\\ & \hspace{0.4cm} - {\widehat Q} \left({\bf s}(t-1),{\bf a}(t)\right) \Big)^2 
\end{align} 
  the Q-function is updated using stochastic gradient descent as 
 \begin{align}
	&\hat{Q}_t\left({\bf s}(t-1),{\bf a}(t)\right) = (1-\beta_t) \hat{Q}_{t-1}\left({\bf s}(t-1),{\bf a}(t)\right)   \nonumber+  \\ 
		\nonumber
		&  \beta_t  \Big[C\left( {{\bf s} (t-1), {\bf a} (t)} {\Big |}  {\mathbf{p}_{\text{G}}(t)},{\mathbf{p}_{\text{L}}(t)} \right) + \gamma \min_{{\boldsymbol \alpha}} \hat{Q}_{t-1}\left({\bf s}(t),{{\boldsymbol \alpha}}\right) \Big] \nonumber 
		\end{align}
 while keeping the rest of the entries in $\hat{Q}_t(\cdot,\cdot)$ unchanged.
 
\begin{algorithm}[t]
	\caption{Caching via Q-learning at CCU}
	\label{alg:QR}
	\begin{algorithmic}
		\State	{\bf Initialize}  $\mathbf{s}(0)$ randomly and $\hat{Q}_0(\mathbf{s,a}) = 0 \; \forall \mathbf{s,a}$ \\
		\For  {$t = 1,2,... $}
		
		\State	{Take action ${\bf a}(t) $ chosen probabilistically by \[{\bf a}(t)  = \left\{
			\begin{array}{ll}
			\arg \min \limits_{\bf a} {\hat Q}_{t-1}\left({\bf s} (t-1),{\bf a}\right) & \textrm{w.p. }\;\; 1-\epsilon_t \\
			\textrm{random } \mathbf{a} \in \mathcal{A}  & \textrm{w.p.} \;\; \; \epsilon_t
			\end{array}
			\right. \]  }
		
		\State ${\bf p}_L (t)$ and ${\bf p}_G (t)$ are revealed based on user requests\\
		\State	Set \hspace{1 cm}${\bf s} (t) = \left[{\bf p}_G^{\top} (t), {\bf p}_L^{\top} (t) , {\bf a}(t)^{\top}\right]^{\top}$
		
		\State	Incur cost $C\Big({ {\bf s} (t-1), {\bf a} (t) {\Big |} {\mathbf{p}_{\text{G}}(t)},{\mathbf{p}_{\text{L}}(t)}}\Big)$ 
		\State	Update \hfill
		\begin{align}
		\hspace{0.6cm}	&\hat{Q}_t\left({\bf s}(t-1),{\bf a}(t)\right) = (1-\beta_t) \hat{Q}_{t-1}\left({\bf s}(t-1),{\bf a}(t)\right)   \nonumber \\ 
		\nonumber
		&\hspace{2cm} +  \beta_t  \Big[C\left( {{\bf s} (t-1), {\bf a} (t)} {\Big |}  {\mathbf{p}_{\text{G}}(t)},{\mathbf{p}_{\text{L}}(t)} \right) \\ &\hspace{4cm} + \gamma \min_{{\boldsymbol \alpha}} \hat{Q}_{t-1}\left({\bf s}(t),{{\boldsymbol \alpha}}\right) \Big] \nonumber 
		\end{align}
		\EndFor
	\end{algorithmic}
	
\end{algorithm}

 Regarding convergence of the Q-learning algorithm, a necessary condition ensuring ${\hat Q}_{t}\left(\cdot,\cdot\right) \rightarrow {Q}^{*}\left(\cdot,\cdot\right)$, is  that  all state-action pairs must be continuously updated \cite{WatkinsQ}. Under this and the usual stochastic approximation conditions that will be specified later, ${\hat Q}_t \left(\cdot,\cdot\right)$ converges to ${Q}^{*} \left(\cdot,\cdot\right)$ with probability~$1$; see~\cite{Stochastic_app_and_reinforcement_learning} for a detailed description.

 {By defining   $t^i(\bf s,\bf a)$ as the index of the $i^{th}$ time that state-action $(\bf s, \bf a)$ is visited, convergence  ${\hat Q}_{t}\left(\cdot,\cdot\right) \rightarrow {Q}^{*}\left(\cdot,\cdot\right)$ can be guaranteed if the stepsize sequence $\{\beta_{t^i(\bf s, \bf a)}\}_{i=1}^\infty$ satisfies   $\sum_{t=1}^{\infty} \{\beta_{t^i(\bf s, \bf a)}\}_{i=1}^\infty = \infty$ and  $\sum_{t=1}^{\infty} \{\beta_{t^i(\bf s, \bf a)}\}_{i=1}^2 < \infty,  \; \forall \bs, \ba$~\cite{WatkinsQ}. 
 Various exploration-exploitation algorithms have been proposed within the scope of \textit{multi-armed bandit problems} \cite{AIModernapproach}, and reasonable schemes have been discussed  that will eventually lead to optimal actions by the agent. Technically, for a constant selection of step size $\beta_t=\beta$, any such scheme needs to be greedy in the limit of infinite exploration, or GLIE \cite[p. 840]{AIModernapproach}. Several GLIE schemes have been proposed, including the $\epsilon_t$-greedy algorithm \cite{wyatt1998exploration} with $\epsilon_t = 1/t$, which will converge to an optimal policy, although at a very slow rate.  Instead, selecting a constant value for  $\epsilon_t$ approaches the optimal $Q^*(.,.)$ faster,  however, since it is not GLIE, its  exact convergence   can not be guaranteed.}
  Additionally, with constant $\epsilon$ as well as  stepsize $\beta_t=\beta$, the mean-square error (MSE) of ${\hat Q}_{t+1}(\cdot,\cdot)$ is bounded as (cf.~\cite{Stochastic_app_and_reinforcement_learning})
 \begin{equation}
 \label{eq_Qhat_Q}
 \mathbb{E} \left[ \left\|{\hat Q_{t+1}}-Q^{*} \right\|_{F}^2 \left| {\hat Q_0} \right. \right] \le \varphi_1\left(\beta\right) +  \varphi_2({\hat Q}_0) \exp\left(-2\beta t\right) 
 \end{equation}
 where $\varphi_1 \left( \beta \right)$ is a positive and increasing function of $\beta$; while the second term denotes the initialization error, which decays exponentially as the iterations proceed.

 {
 Current work utilizes an $\epsilon_t$-greedy exploration-exploitation approach to selecting actions. 	 
To this end, during initial iterations or when the CCU observes considerable shifts in content popularities, setting $\epsilon_t$ high promotes exploration in order to learn the underlying dynamics. On the other hand, in stationary settings and once ``enough'' observations are made, small values of $\epsilon_t$ is desirable as it results agent actions to approach the optimal policy.} 

Although selection of a constant stepsize prevents the algorithm from exact convergence to  $Q^*$ in stationary settings, it enables CCU adaptation to the underlying non-stationary Markov processes in dynamic scenaria. Furthermore, the optimal policy in practice can be obtained from the Q-function values before convergence is achieved \cite[pg. 79]{Sutton1998reinforcement}.

However, the main practical limitation of the Q-learning algorithm is its slow convergence, which is a consequence of independent updates of the Q-function values. Indeed, Q-function values are related, and leveraging these relationships can lead to multiple updates per observation as well as faster convergence. In the ensuing section, the structure of the problem at hand will be exploited  to develop a linear function approximation of the Q-function, which in turn  will endow our algorithm not only with fast convergence, but also with scalability.

\section{Scalable caching}
\label{Linear Function approximation}
Despite simplicity of the updates as well as optimality guarantees of the Q-learning algorithm, its applicability over real networks faces practical challenges. Specifically, the Q-table is of size $|\mathcal{P}_G| |\mathcal{P}_L| |\mathcal{A}|^2$, where  $|\mathcal{A}|={F \choose M} $ encompasses all possible selections of $M$ from $F$ files. Thus, the Q-table size grows prohibitively with $F$, rendering convergence of the table entries, as well as the policy iterates unacceptably slow. Furthermore, action selection in $\min_{{\boldsymbol \alpha} \in {\cal A}} Q({\bf s},{\bf a})$ entails an expensive exhaustive search over the feasible action set $\mathcal{A}$. 

Linear function approximation is  a popular scheme for rendering Q-learning applicable to real-world settings \cite{geramifard2013tutorial,mahadevan2009learning,AIModernapproach}.  
A linear approximation for $Q(\mathbf{s},\mathbf{a})$ in our setup is inspired by the additive form of the instantaneous costs in \eqref{Overall_Cost}. Specifically, we propose to approximate $Q(\mathbf{s},\mathbf{a}')$ as  
\begin{equation}
\label{approximation}
Q(\mathbf{s},\mathbf{a}')\simeq Q_G(\mathbf{s},\mathbf{a}')+Q_L(\mathbf{s},\mathbf{a}')+Q_R(\mathbf{s},\mathbf{a}')
\end{equation}
where $Q_G$, $Q_L$, and $Q_R$ correspond to global and local popularity mismatch, and cache-refreshing costs, respectively.

Recall that the state vector $\mathbf{s}$ consists of three subvectors, namely $\mathbf{s} := [{\mathbf p}^{\top}_G, {\mathbf p}^{\top}_L,\mathbf{a}^{\top}]^{\top}$. Corresponding to the global popularity subvector, our first term of the approximation in \eqref{approximation} is 
\begin{align}
\label{Q_G_est_1}
Q_G(\mathbf{s},\mathbf{a}'):= \sum_{i=1}^{|\mathcal{P}_G|}\sum_{f=1}^F \theta^{G}_{i,f} \mathbbm{1}_{\left\{\mathbf{p}_G={\bf p}^{i}_{G}\right\}} \mathbbm{1}_{\left\{[{\ba}']_f=0\right\}}
\end{align}	
where the sums are over all possible global popularity profiles as well as  files, and the indicator function ${\mathbbm{1}}_{\left\{ \cdot \right\}}$ takes value $1$ if its argument holds, and $0$ otherwise; while $\theta^{G}_{i,f}$ captures the average ``overall'' cost if the system is in global state ${\bf p}^{i}_G$, and the CCU decides not to cache the $f$th content. By defining the  $|{\cal P}_G| \times |\cal F|$ matrix with $(i,f)$-th entry $\left[{\boldsymbol \Theta}^G\right]_{i,f} := \theta^{G}_{i,f}$, one can rewrite \eqref{Q_G_est_1} as 
\begin{equation}
\label{Matrix_G}
Q_G(\mathbf{s},\mathbf{a}')= \boldsymbol{\delta}_G^{\top}({\bp_G}) \boldsymbol{\Theta}^G (\mathbf{1}-{\bf a}')
\end{equation}  
where 
\[ {\boldsymbol \delta}_G ({\bf p}_G) := \left[\delta({\bf p}_G-{\bf p}^{1}_{G}), \ldots, \delta({\bf p}_G-{\bf p}^{|{\cal P}_G|}_{G})\right]^{\top}.\]

Similarly, we advocated the second summand in the approximation  \eqref{approximation} to be
\begin{align}
\nonumber
Q_L(\mathbf{s},\mathbf{a}')&:= \sum_{i=1}^{|\mathcal{P}_L|}\sum_{f=1}^F \theta^{L}_{i,f} \mathbbm{1}_{\left\{\mathbf{p}_L={\bf p}^{i}_{L}\right\}} \mathbbm{1}_{\left\{[{\ba}']_f=0\right\}} \\ \label{Q_L_est_2} &   = \boldsymbol{\delta}^\top_{L} ({{\bp}_L}) \boldsymbol{\Theta}^L (\mathbf{1}-{\bf a}')
\end{align}
where $\left[{\boldsymbol \Theta}^L\right]_{i,f} := \theta^{L}_{i,f}$, and  
\[ {\boldsymbol \delta}_L ({\bf p}_L) := \left[\delta({\bf p}_L-{\bf p}^{1}_{L}), \ldots, \delta({\bf p}_L-{\bf p}^{|{\cal P}_L|}_{L})\right]^{\top}\]
with $\theta^{L}_{i,f}$ modeling the average overall cost for not caching file $f$ when the local popularity is in state ${\bf p}^i_L$.

Finally, our third summand in \eqref{approximation} corresponds to the cache-refreshing cost
\begin{align}
Q_R(\mathbf{s},\mathbf{a}'):&= \sum_{f=1}^F \theta^{R} \mathbbm{1}_{\left\{[{\ba}']_f=1\right\}} \mathbbm{1}_{\left\{[\ba]_f=0\right\}} \\ & = \theta^{R} {\bf a}'^{\top} \left(1-{\bf a}\right)  \nonumber \\ &= \theta^{R} \left[ {\bf a}'^{\top} \left(1-{\bf a}\right) +{\bf a}^{\top} {\bf 1} - {\bf a}'^{\top} {\bf 1}\right] \nonumber
 \\ &   = \theta^{R} {\bf a}^{\top} (\mathbf{1}-{\bf a}') \nonumber 
\label{eq.Qr}
\end{align}
where $\theta^R$ models average cache-refreshing cost per content. The constraint {$\mathbf{a}^\top \mathbf{1}= \mathbf{a}'^\top \mathbf{1} = M$},
is utilized to factor out the term $\bf{1-a}'$, which will become useful later.

Upon defining the set of parameters  $\Lambda := \{\boldsymbol{\Theta}^G,\boldsymbol{\Theta}^L, \theta^a\}$, the Q-function is readily approximated (cf. \eqref{approximation}) 
\begin{equation}
\label{eq.app}
{\widehat Q}_{\Lambda}(\bs,{\ba}'):= \underbrace{\Big( \boldsymbol{\delta}_G^{\top}({\bp_G}) \boldsymbol{\Theta}^G +  \boldsymbol{\delta}_L^{\top}({\bp_L}) \boldsymbol{\Theta}^L    + \theta^{R} {\bf a}^{\top} \Big)}_{\boldsymbol \psi(\bf s):=} (\mathbf{1}-{{\ba}'}).
 \end{equation}
 Thus, the original task of learning $|{\cal P}_G| |{\cal P}_L| |{\cal A}|^2$ parameters  in Alg.~1 is now reduced to learning $\Lambda$ containing $\left( \left|{\cal P}_G\right| + \left|{\cal P}_L\right| \right) \left|{\cal F}\right|  +1$ parameters.

\subsection{Learning $\Lambda$}
Given the current parameter estimates  $\{\widehat{\boldsymbol{\Theta}}_{t-1}^G,\widehat{\boldsymbol{\Theta}}_{t-1}^L, \hat{\theta}_{t-1}^R\}$ at the end of information exchange phase of slot $t$, the instantaneous error is given by {
\begin{align} \nonumber 
\widehat{e} & \left({\bf s}(t-1), {\bf a}(t)\right)  :=   C\left({\bf s}(t-1),{\bf a}(t)\right) \\ \nonumber &\hspace{+.7cm } + \gamma \min \limits_{{\bf a}'}^{} {\widehat Q}_{{\Lambda_{t-1}}} \left({\bf s}(t), {\bf a}'\right) - {\widehat Q}_{{\Lambda_{t-1}}} \left({\bf s}(t-1),{\bf a}(t)\right) . 
\end{align} 
Let us define 
\begin{equation}
\label{error}
\widehat\varepsilon \left({\bf s}(t-1), {\bf a}(t)\right) := \dfrac{1}{2} \Big(\widehat{e}\left({\bf s}(t-1), {\bf a}(t)\right)\Big)^2,
\end{equation}
then,  the parameter update rules are obtained using stochastic gradient descent iterations as (cf. \cite[p. 847]{AIModernapproach})
\begin{align}
\label{UpdatethetaG}
& \hat{\boldsymbol{\Theta}}_t^G = \hat{\boldsymbol{\Theta}}_{t-1}^G - \alpha_G \nabla_{{\boldsymbol{\Theta}}^G} \widehat\varepsilon \left({\bf s}(t-1), {\bf a}(t)\right) 
\\ & = \hat{\boldsymbol{\Theta}}_{t-1}^G + \alpha_G \;{\widehat{e}  \left({\bf s}(t-1), {\bf a}(t)\right) } \; \;  \nabla_{{\boldsymbol \Theta}^{G}} {\widehat Q}_{{\Lambda_{t-1}}} ({\bf s}(t-1),{\bf a}(t))\nonumber \\  &
= \hat{\boldsymbol{\Theta}}_{t-1}^G + \alpha_G \;{\widehat{e}  \left({\bf s}(t-1), {\bf a}(t)\right) } \; \;   \boldsymbol{\delta }_G ({\bp_G(t-1)}) (\mathbf{1}-\ba(t))^\top \nonumber
\end{align} 
\begin{align}
\label{UpdatethetaL}
& \hat{\boldsymbol{\Theta}}_t^L = \hat{\boldsymbol{\Theta}}_{t-1}^L - \alpha_L \nabla_{{\boldsymbol{\Theta}}^L} \widehat\varepsilon \left({\bf s}(t-1), {\bf a}(t)\right) 
\\ & = \hat{\boldsymbol{\Theta}}_{t-1}^L + \alpha_L \;{\widehat{e}  \left({\bf s}(t-1), {\bf a}(t)\right) } \; \;  \nabla_{{\boldsymbol \Theta}^{L}} {\widehat Q}_{{\Lambda_{t-1}}} ({\bf s}(t-1),{\bf a}(t))\nonumber \\  
& = \hat{\boldsymbol{\Theta}}_{t-1}^L + \alpha_L \;{\widehat{e}  \left({\bf s}(t-1), {\bf a}(t)\right) } \; \;   \boldsymbol{\delta }_L ({\bp_L(t-1)}) (\mathbf{1}-\ba(t))^\top \nonumber
\end{align} 
and 
\begin{align}
\label{UpdatethetaR}
& \hat{{\theta}}_t^R = \hat{{\theta}}_{t-1}^R - \alpha_R \nabla_{{{\theta}}^R} \widehat\varepsilon \left({\bf s}(t-1), {\bf a}(t)\right)
\\ &= \hat{{\theta}}_{t-1}^R + \alpha_R \;{\widehat{e}  \left({\bf s}(t-1), {\bf a}(t)\right) } \; \;  \nabla_{{\theta}^R} {\widehat Q}_{\Lambda_{t-1}} ({\bf s}(t-1),{\bf a}(t))\nonumber \\  
&= \hat{{\theta}}_{t-1}^R + \alpha_R \;{\widehat{e}  \left({\bf s}(t-1), {\bf a}(t)\right) } \; \;  {\bf a}^{\top}(t-1) (\mathbf{1}-{\bf a}(t)) \nonumber.
\end{align}
}
The pseudocode for this scalable approximation of the Q-learning scheme is tabulated in Alg.~2. 
\begin{algorithm}[t]
	\caption{Scalable Q-learning}
	\label{LFAQ}
	\begin{algorithmic}
		\State	{\bf Initialize}  $\mathbf{s}(0)$ randomly, ${\widehat{\boldsymbol \Theta}_0^G} = {\bf 0}$, ${\widehat{\boldsymbol \Theta}_0^L} = {\bf 0}$, ${\hat \theta}_0^R = 0$, and thus $\widehat{\boldsymbol{\psi}}(\bf s) = {\bf 0}$ \\
		\For  {$t=1,2,...$}
		
		\State	{Take action ${\bf a}(t) $ chosen probabilistically by \[{\bf a}(t)  = \left\{
			\begin{array}{ll}
			\text {$M$ best files via ${\widehat{\boldsymbol \psi} \left({\bf s}(t-1)\right)}$} & \textrm{w.p. }\;\; 1-\epsilon_t \\
			\textrm{random } \mathbf{a} \in \mathcal{A}  & \textrm{w.p.} \;\; \; \epsilon_t
			\end{array}
			\right. \]  \hspace{0.4 cm} where 	 ${\boldsymbol{ \hat{\psi}} ({\bf s})} := \boldsymbol{\delta}_G^{\top}({\bp_G}) \boldsymbol{\widehat{\Theta}}^G +  \boldsymbol{\delta}_L^{\top}({\bp_L}) \boldsymbol{\widehat{\Theta}}^L    + \widehat{\theta}^{R} {\bf a}^{\top}$}
		\vspace{0.3cm}
		\State ${\bf p}_G (t)$ and ${\bf p}_L (t)$ are revealed based on user requests\\
		\State	Set \quad  \hspace{0.9 cm} ${\bf s} (t) = \left[{\bf p}_G^{\top} (t), {\bf p}_L^{\top} (t) , {\bf a}(t)^{\top}\right]^{\top}$
		
		\State	Incur cost \quad $C\Big({ {\bf s} (t-1), {\bf a} (t) {\Big |} {\mathbf{p}_{\text{G}}(t)},{\mathbf{p}_{\text{L}}(t)}}\Big)$ 
		\State Find \hspace{1 cm} $\widehat \varepsilon \left({\bf s}(t-1),{\bf a}(t)\right)$
		\State	Update \quad \hspace{0.3cm} ${\widehat{\boldsymbol{\Theta}}}_t^G$, ${\widehat{\boldsymbol{\Theta}}}_t^L$ and ${\hat \theta}_t^R$ based on \eqref{UpdatethetaG}-\eqref{UpdatethetaR}
		\EndFor
	\end{algorithmic}
	
\end{algorithm}                
The upshot of this scalable scheme is three-fold. 
\begin{itemize}
	\item The large state-action  space in the Q-learning algorithm is handled by reducing the number of parameters from $|{\cal P}_G| |{\cal P}_L| |{\cal A}|^2$ to  $\left( \left|{\cal P}_G\right| + \left|{\cal P}_L\right| \right) \left|{\cal F}\right| +1$.
	\item In contrast to single-entry updates in the exact Q-learning Alg.~1, $F-M$ entries in  $\widehat{\boldsymbol\Theta}^G$ and $\widehat{\boldsymbol\Theta}^L$ as well as $\theta^R$, are updated per observation using \eqref{UpdatethetaG}-\eqref{UpdatethetaR}, which leads to a much faster convergence.
	\item The exhaustive search in $ \min \limits_{{\bf a} \in {\cal A}} Q \left({\bf s},{\bf a}\right)$ required in exploitation; and also in the error evaluation \eqref{error}, is circumvented. Specifically, it holds that (cf.~\eqref{eq.app}) \begin{equation}
	\label{a}
	\min \limits_{{\bf a}' \in {\cal A}} Q({\bf s},{\bf a}') \approx \min \limits_{{\bf a}' \in {\cal A}} {\boldsymbol \psi}^{\top} ({\bf s}) \left({\bf 1- a}' \right) = \max  \limits_{{\bf a}'\in {\cal A}} {\boldsymbol \psi}^{\top}({\bf s}) \, \, {\bf a}'
	\end{equation}
	where \[{\boldsymbol{ \psi} ({\bf s})} := \boldsymbol{\delta}_G^{\top}({\bp_G}) \boldsymbol{\Theta}^G +  \boldsymbol{\delta}_L^{\top}({\bp_L}) \boldsymbol{\Theta}^L    + \theta^{R} {\bf a}^{\top}. \] 
	The solution of \eqref{a} is readily given by $[{\bf a}]_{\nu_i} = 1$ \linebreak for $i=1, \ldots, M$, and $[{\bf a}]_{\nu_i} = 0$ for $i>M$, where $\left[{{\boldsymbol{ \psi} ({\bf s})}}\right]_{\nu_F} \le \cdots \le  \left[{{\boldsymbol{ \psi} ({\bf s})}}\right]_{\nu_1} $ are sorted entries of ${\boldsymbol{ \psi} ({\bf s})}$.  
\end{itemize}

\noindent{\bf Remark 2.}
In the model of Sec. II-B, the state-space cardinality of the popularity vectors is finite. These vectors can be viewed as centroids of quantization regions partitioning a state space of infinite cardinality. Clearly, such a partitioning  inherently bears a complexity-accuracy trade off, motivating  optimal designs to achieve a desirable accuracy for a given affordable complexity. This is one of our future research directions for the  problem at hand.

Simulation based evaluation  of the proposed algorithms for RL-based caching is now in order.


\section{Numerical tests}

\begin{figure}[t]
	\centering
	\begin{subfigure}{0.4\textwidth}
		\centering
		\includegraphics[width=0.82\columnwidth]{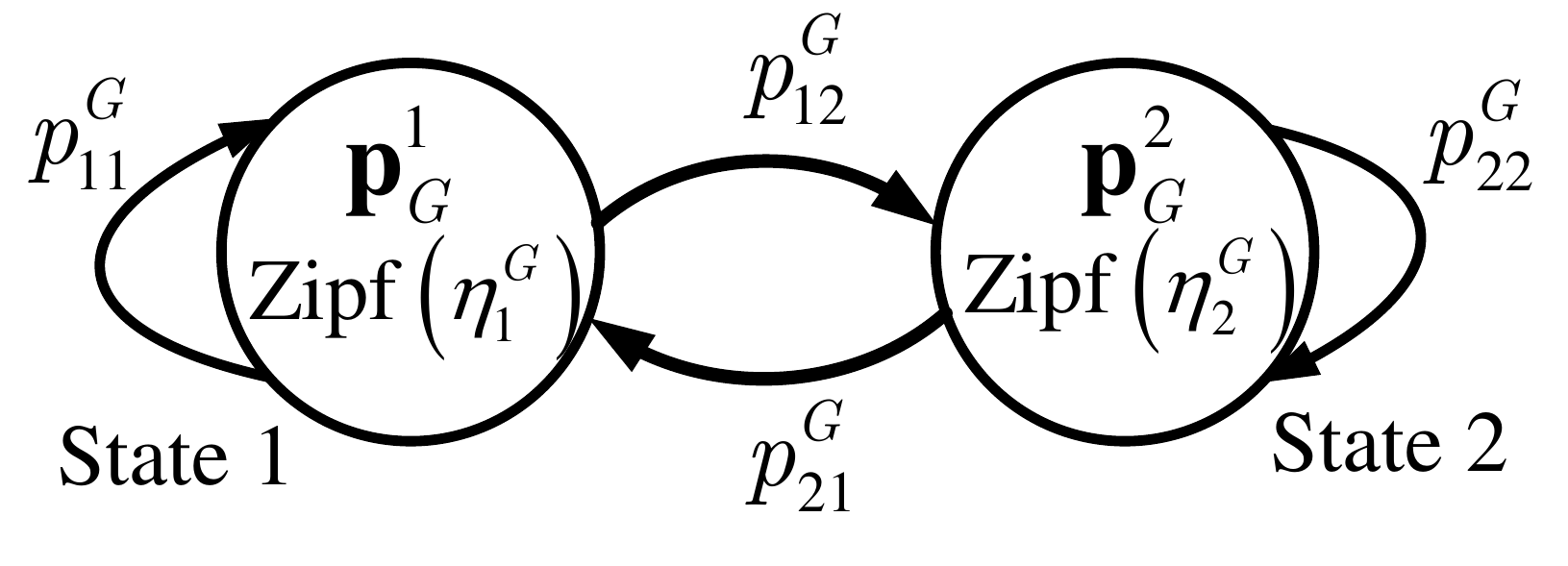}
		\caption{Global popularity Markov chain.}
	\end{subfigure} \\
	\begin{subfigure}{0.4\textwidth}
		\centering
		\includegraphics[width=0.82\columnwidth]{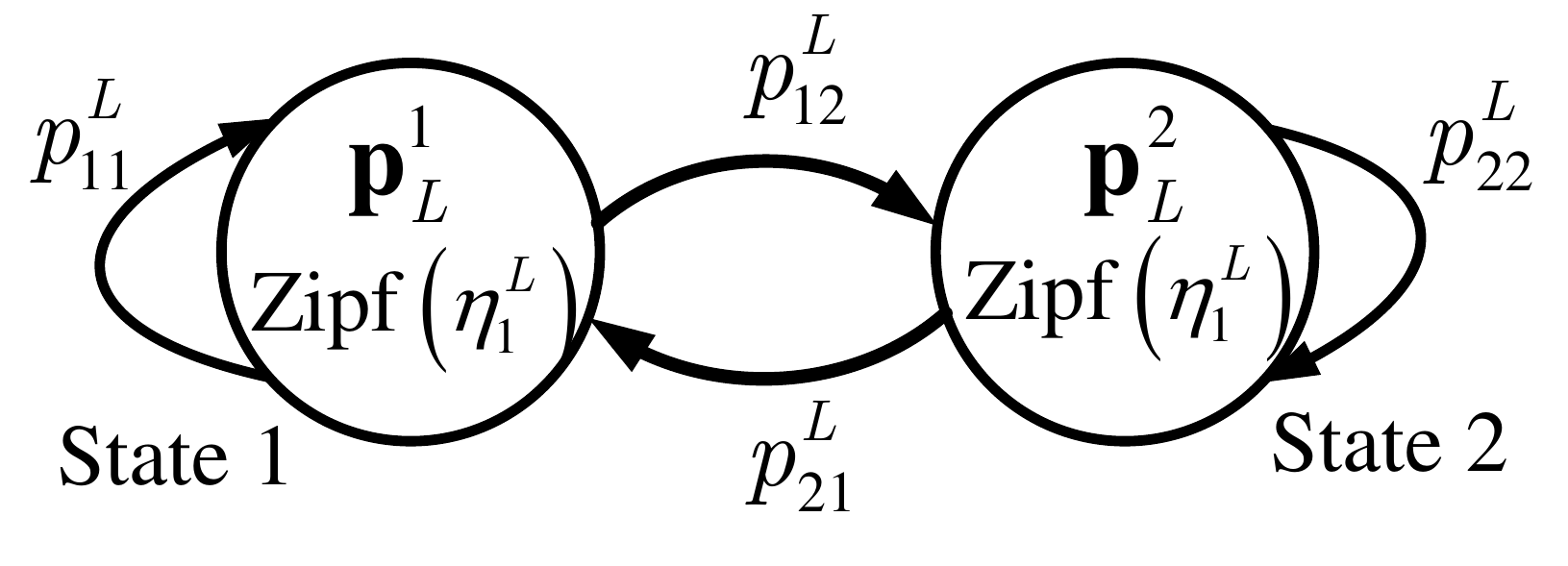}
		\caption{Local popularity Markov chain.}
	\end{subfigure}
	\caption{Popularity profiles Markov chains.}
	\label{Markovchains}
\end{figure}

In this section, performance of the proposed Q-learning algorithm and its scalable approximation is tested. To compare the proposed algorithms with the optimal caching policy, {which is the best policy under \textit{known} transition probabilities for global and local popularity Markov chains}, we first simulated a small network with $F = 10$ contents, and caching capacity $M = 2$ at the local SB. Global popularity profile is modeled by a two-state Markov chain with states ${\bf p}^{1}_G$ and
${\bf p}^{2}_G$, that are drawn from Zipf distributions having parameters $\eta_1^G=1$ and $\eta_2^G=1.5$, respectively \cite{breslau1999web}; see also Fig.~\ref{Markovchains}.   That is, for state $i \in \left\{1,2\right\}$, the $F$ contents are assigned a random ordering of popularities, and then sorted accordingly in a descending order. Given this ordering and the Zipf distribution parameter $\eta_i^G$, the popularity of the $f$-th  content is set to  
\begin{equation}
\nonumber
\bigg[\mathbf{p}_{\text{G}}^{i} \bigg]_f = \frac{1}{f^{{\eta^G_i}} \sum \limits_{l=1}^{F}   1 \mathbin{/} l^{\eta_i^G}} \;\quad \text{{for}} \;\;i=1,2 \end{equation} 
where the summation normalizes the components to follow a valid probability mass function, while $\eta_i^G \geq 0$ controls the skewness of popularities. Specifically, $\eta_i^G=0$ yields a  uniform spread of popularity among contents, while a large value of $\eta_i$ generates more skewed popularities. 
Furthermore, state transition probabilities  of the Markov chain modeling  global popularity profiles are 
\begin{equation}
\nonumber
{\bm P}^{G}:= \left[ {\begin{array}{*{20}{c}}
	{{p^{G}_{11}}}&{{p^{G}_{12}}}\\{{p^{G}_{21}}}&{{p^{G}_{22}}}
	\end{array}} \right] = \left[ {\begin{array}{*{20}{c}}
	{{0.8}}&{{0.2}}\\ {0.75}&{0.25}
	\end{array}} \right].
\end{equation} 

Similarly, local popularities are modeled by a two-state Markov chain, with states $\mathbf{p}_L^{1}$ and  $\mathbf{p}_L^{2}$, whose entries are drawn from Zipf distributions with parameters $\eta^{L}_1 = 0.7$ and $\eta^{L}_2 = 2.5$, respectively. The transition probabilites of the local popularity Markov chain are  
\begin{equation}
\nonumber
{\bm P}^{L}:= \left[ {\begin{array}{*{20}{c}}
	{{p^{L}_{11}}}&{{p^{L}_{12}}}\\{{p^{L}_{21}}}&{{p^{L}_{22}}}
	\end{array}} \right] = \left[ {\begin{array}{*{20}{c}}
	{{0.6}}&{{0.4}}\\ {0.2}&{0.8}
	\end{array}} \right].
\end{equation}

Caching performance is assessed under {three} cost-parameter settings: (s1)   $\lambda_1 = 10, \lambda_2 = 600, \lambda_3 = 1000$; (s2) $\lambda_1 = 600, \lambda_2 = 10, \lambda_3 = 1000$, { and (s3) $\lambda_1 = 10, \lambda_2 = 10, \lambda_3 = 1000$}. {In all numerical tests the optimal caching policy is found by utilizing the policy iteration algorithm with known transition probabilities.} In addition, Q-learning in Alg. 1 and its scalable approximation in Alg. 2 are run with $\beta_t = 0.8$, $\alpha_{G} = \alpha_{L} = \alpha_{R} = 0.005$, and $\epsilon_t = 0.05$.

\begin{figure}[t]
	\centering
	\includegraphics[width=1\columnwidth]{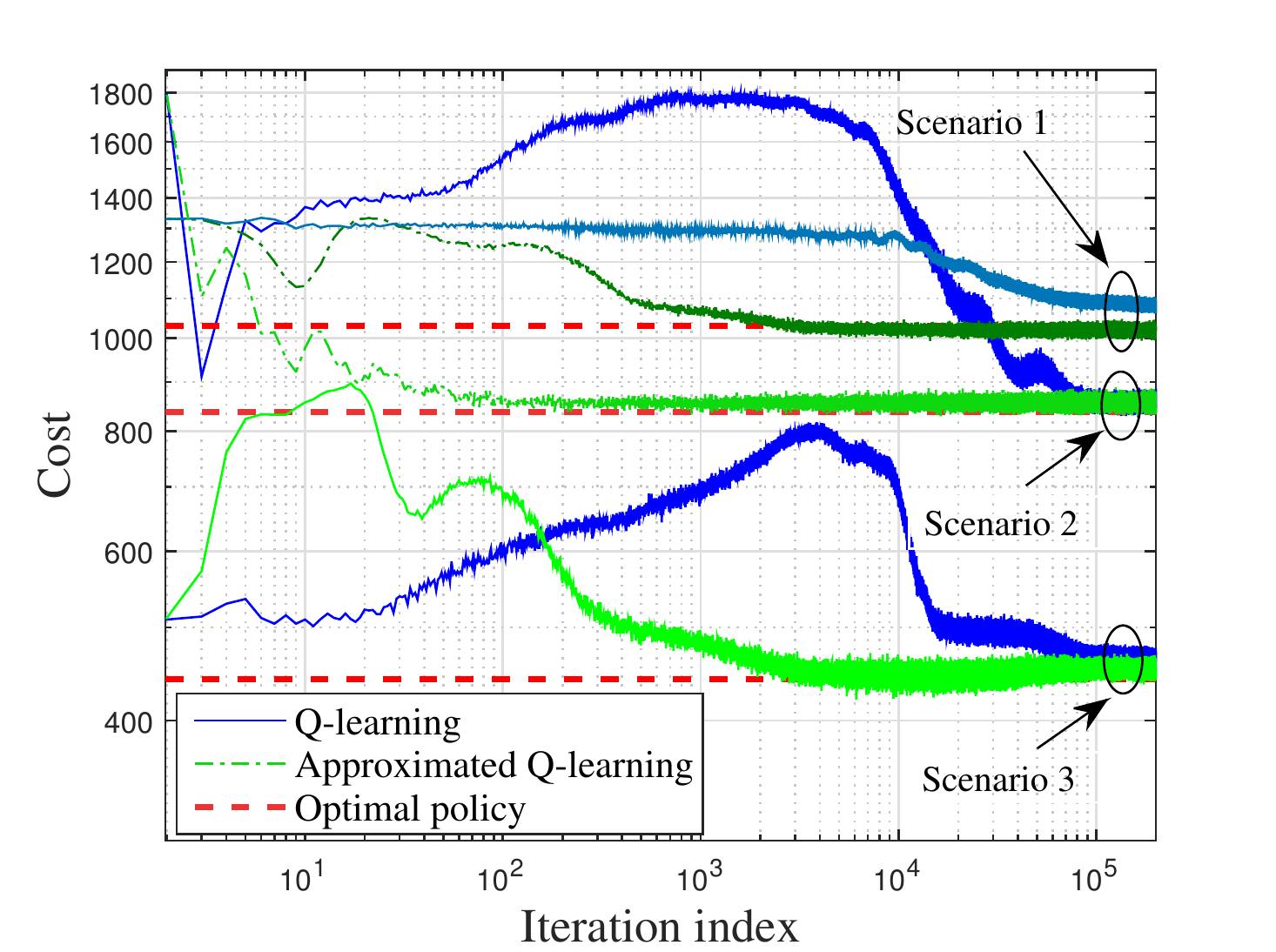}
	\caption{Performance of the proposed algorithms.}
	\label{Cost}
\end{figure}

Fig.~\ref{Cost} depicts the observed cost versus iteration (time) index averaged over 1000 realizations. It is seen that the cashing cost via Q-learning, and through its scalable approximation converge to that of the  optimal  policy. As anticipated,  even for the small size of this network, namely $|{\cal P}_G| = |{\cal P}_L| = 2$ and $|{\cal A}| = 45$, the Q-learning algorithm converges slowly to the optimal policy, especially under s1, while its scalable approximation exhibits faster convergence. {The reason for slower convergence under (s1) is that the corresponding cost parameters of local and global popularity mismatch are set  high, thus, the convergence of the Q-learning algorithm as well as caching policy essentially relies on learning both global and local popularity Markov chains. In contrast, under (s2), $\lambda_2$ corresponding to  local popularity mismatch is low, thus the impact  of local popularity Markov chain on the optimal policy is reduced, giving rise to a simpler policy, thus a faster convergence.  To further elaborate this issue, simulations are carried out under a simpler scenario (s3). In this setting, having $\lambda_1 = 10$ further reduces the effect of cache refreshing cost, and thus more importance falls on learning the Markov chain of global popularities. Indeed, the simulations present a slightly faster convergence for (s3) compared to (s2), while both demonstrate  much faster convergence than (s1).}

\begin{figure}[t]
	\centering
	\includegraphics[width=1\columnwidth]{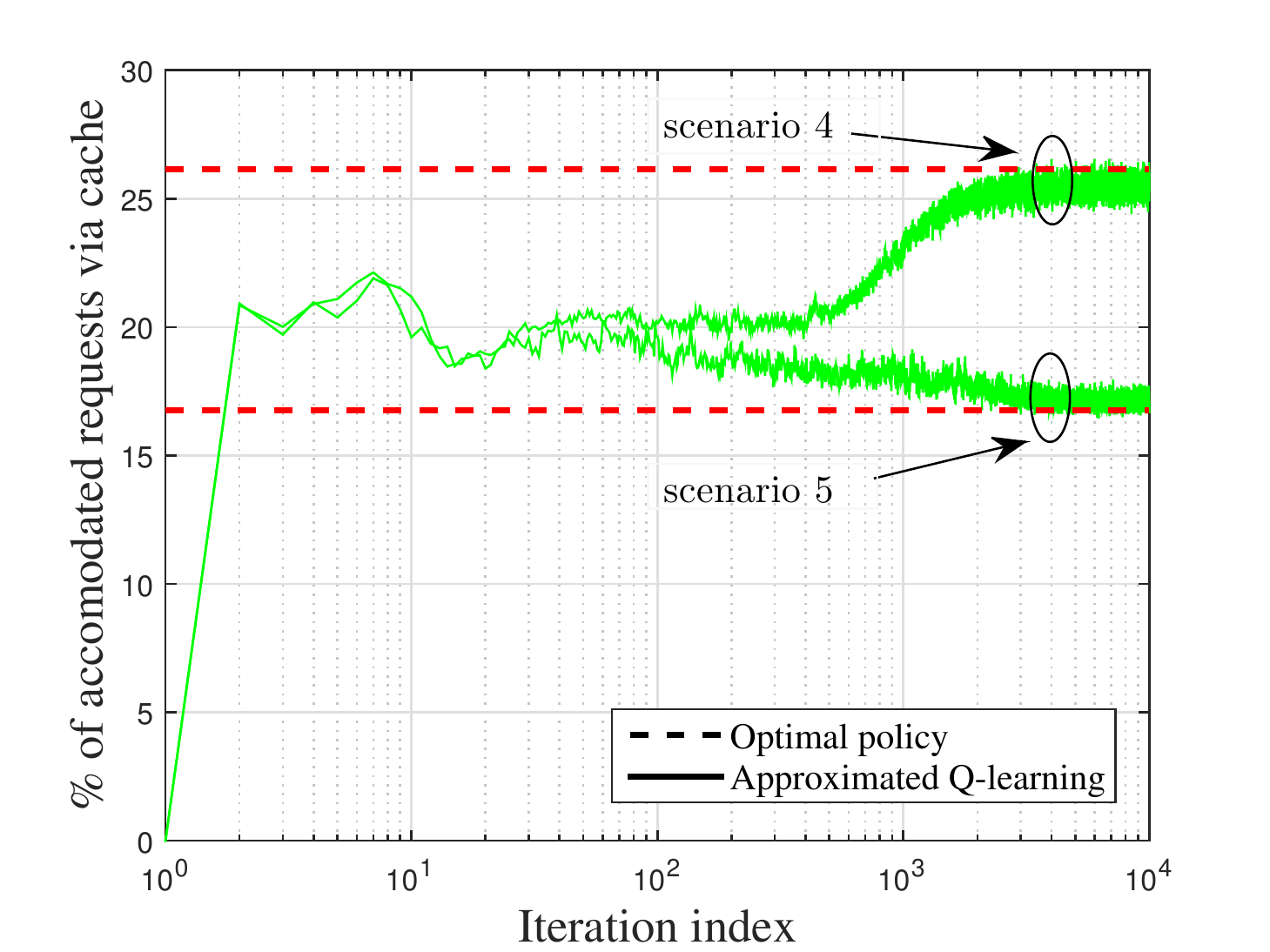}
	\caption{Percentage of accommodated requests via cache.}
	\label{percent}
\end{figure}

In order to highlight the trade-off between global and local popularity mismatches, 
the percentage of accommodated requests  via cache is depicted in Fig. \ref{percent} for settings (s4)  $\lambda_1 = \lambda_3 = 0, \lambda_2 = 1,000$, and (s5) $ \lambda_1 = \lambda_2 = 0, \lambda_3 = 1,000$. 
Observe that penalizing local popularity-mismatch in (s4) forces  the caching policy to  adapt to local request dynamics, thus accommodating a higher percentage of requests via cache, while (s5) prioritizes tracking global popularities, leading  to a lower cache-hit in this setting. Due to slow convergence of the exact Q-learning under (s4) and (s5), only the performance of the scalable solver is presented here. 


\begin{figure}[t]
	\centering
	\includegraphics[width=1\columnwidth]{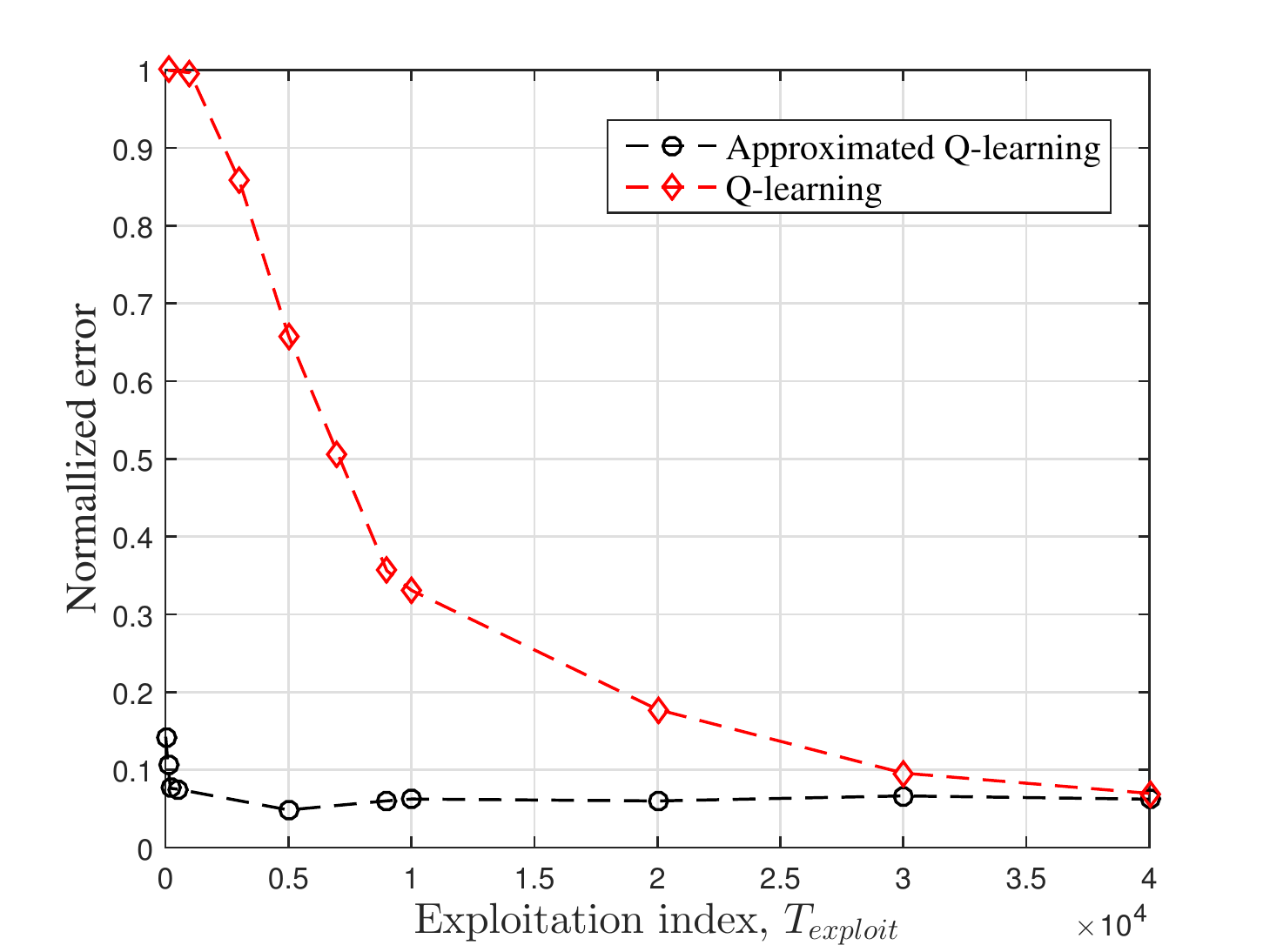}
	\caption{Convergence rate of the exact and scalable Q-learning.}
	\label{Convergence}
\end{figure}

Furthermore, the convergence rate of Algs.~1 and 2 is illustrated under {(s6) $ \lambda_1 = 60, \lambda_2 = 10, \lambda_3 = 10$} in Fig.~\ref{Convergence}, where average normalized error is evaluated in terms of the ``exploitation index.'' Specifically,  a pure exploration is taken for the first $T_{\rm{explore}}$ iterations of the algorithms, i.e., $\epsilon_t = 1$ for $t=1,2,\ldots,T_{\rm{explore}}$; and a pure exploitation  with $\epsilon_t = 0$ is adopted afterwards. 
We have set $\alpha=0.005$, and selected  $\beta_t=0.7$. As the plot demonstrates, the exact Q-learning Alg.~1 exhibits slower convergence, whereas just a few iterations suffice for the scalable Alg. 2 to converge to the optimal solution, thanks to the reduced dimension of the problem as well as the multiple updates that can be afforded per iteration.

Having established the accuracy and efficiency  of the Alg. 2, we next simulated a larger network with $F = 1,000$ available files, and a cache capacity of $M = 10$, giving rise to a total of ${{1000}\choose{10} }\simeq 2 \times 10^{23}$ feasible caching actions. In addition, we set the local and global popularity Markov chains to have $|{\cal P}_L| = 40$  and $|{\cal P}_G| = 50$ states, for which the underlying state transition probabilities  are drawn randomly, and  Zipf parameters are drawn uniformly over the interval $(2,4)$. 

\begin{figure}[t]
	\centering
	\includegraphics[width=1\columnwidth]{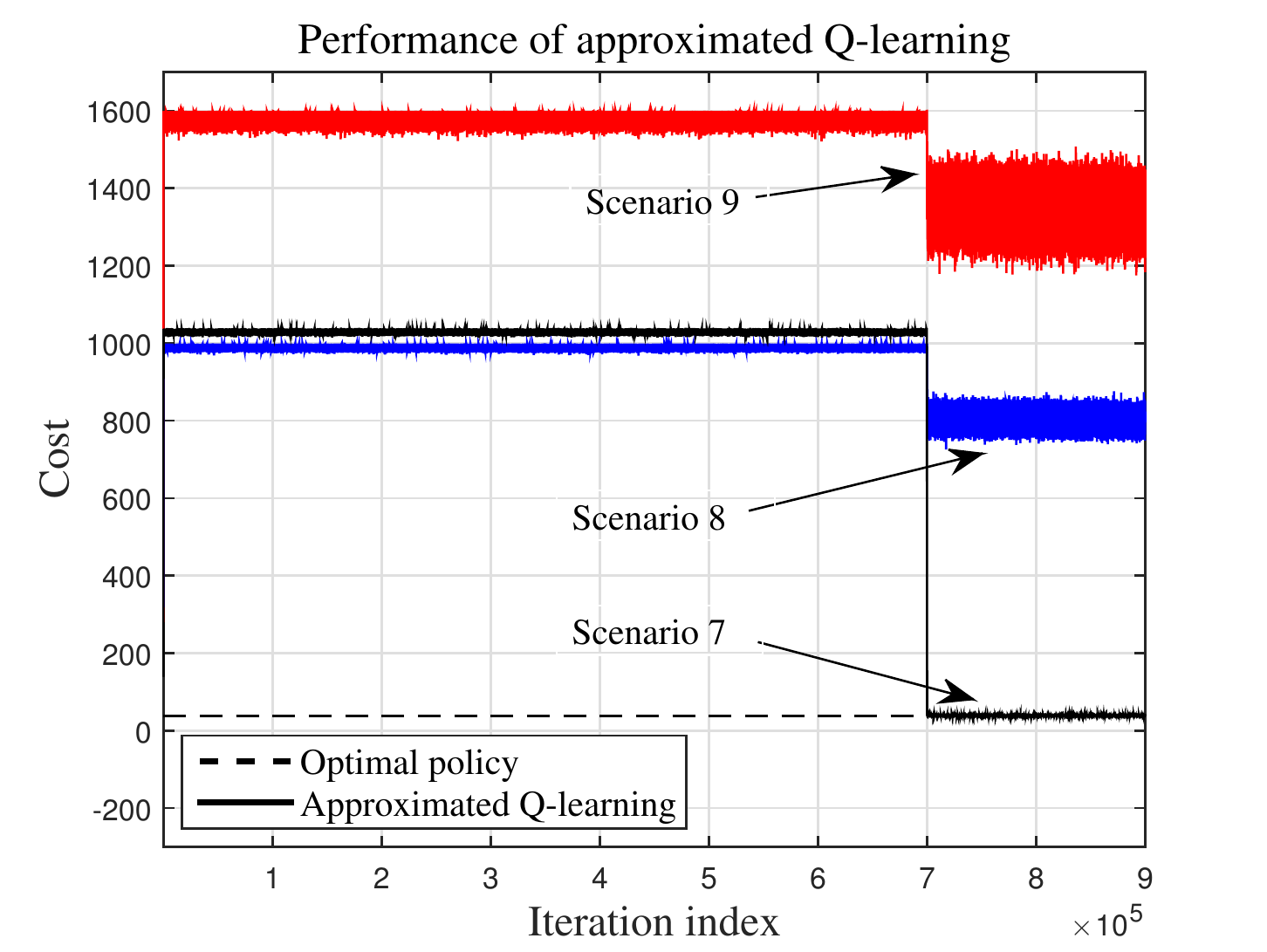}
	\caption{Performance  in large state-action space scenaria.}
	\label{appr}
\end{figure}

Fig.~\ref{appr} plots the performance of Alg. 2 under (s7) $\lambda_1 = 100$, $\lambda_2 =20$, $\lambda_3 =20$, (s8) $\lambda_1 = 0$, $\lambda_2 = 0$, $\lambda_3 = 1,000$, and (s9)~$\lambda_1 = 0$, $\lambda_2 = 1,000$, $\lambda_3 = 600$. Exploration-exploitation parameter is set to  $\epsilon_t=1 $ for $t=1,2,\ldots,7 \times 10^{5}$, in order to greedily explore the entire state-action space in initial iterations, and $\epsilon_t = 1 \mathbin{/} ({\textrm{iteration index}})$ for $t>7 \times 10^{5}$.  

Finding the optimal policy in (s8) and (s9) requires prohibitively sizable memory as well as extremely high computational complexity, and it is thus unaffordable for this network. However, having large cache-refreshing cost with $\lambda_1 \gg \lambda_2,\lambda_3$ in (s7) forces the optimal caching policy to freeze its cache contents, making the optimal caching policy predictable in this setting. Despite the very limited storage capacity, of $10\mathbin{/}1,000=0.01$ of available files, utilization of RL-enabled caching offers a considerable reduction in incurred costs, while the proposed approximated Q-learning endows the approach with scalability and light-weight updates.

\begin{figure}[t]
	\label{profile}
	\centering
	\begin{subfigure}{.5\textwidth}
		\centering
		\includegraphics[width=1\columnwidth]{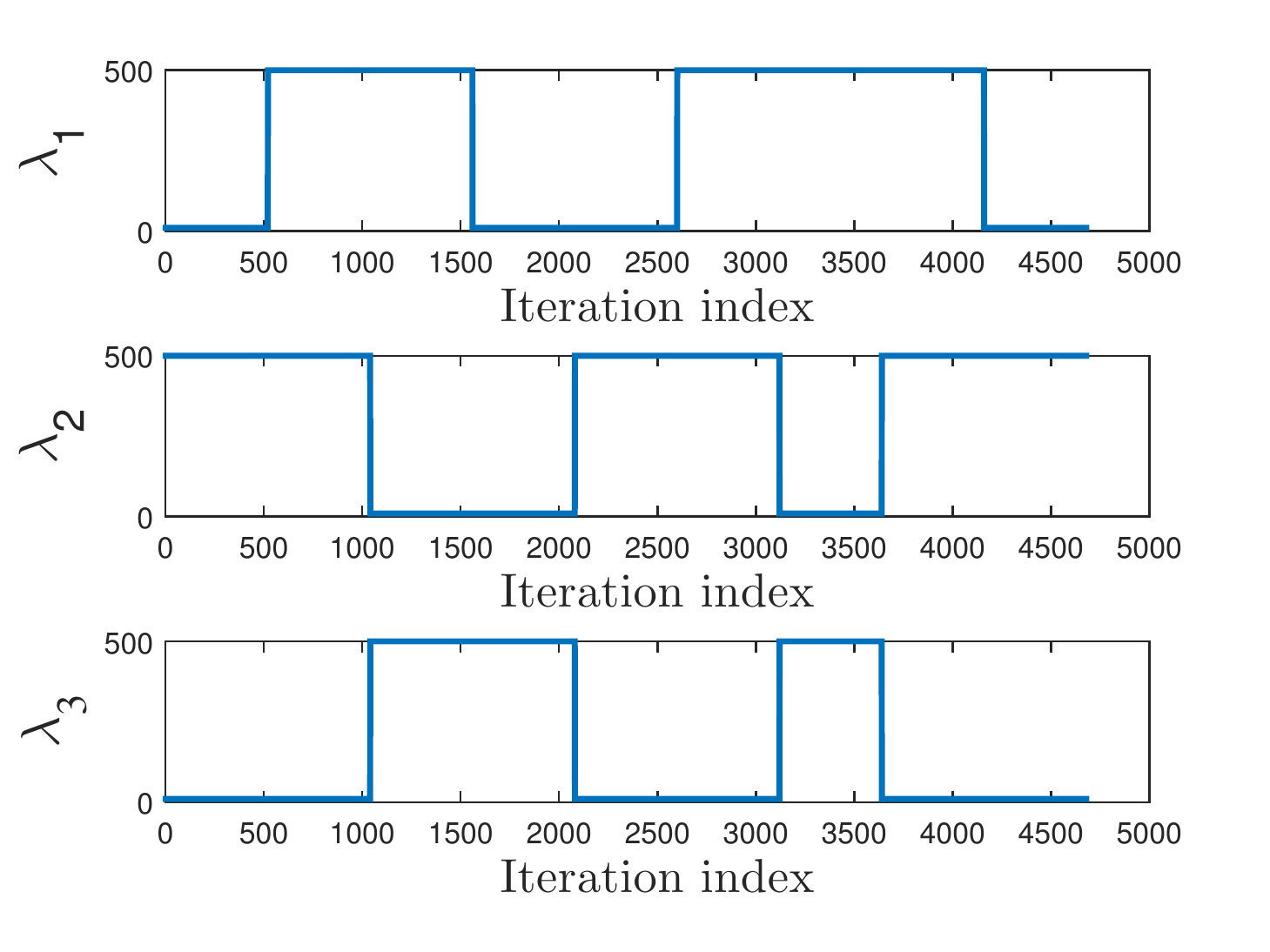}
		\caption{Cost profiles.}
	\end{subfigure}  \\ \begin{subfigure}{0.5\textwidth}
		\centering
		\includegraphics[width=1\columnwidth]{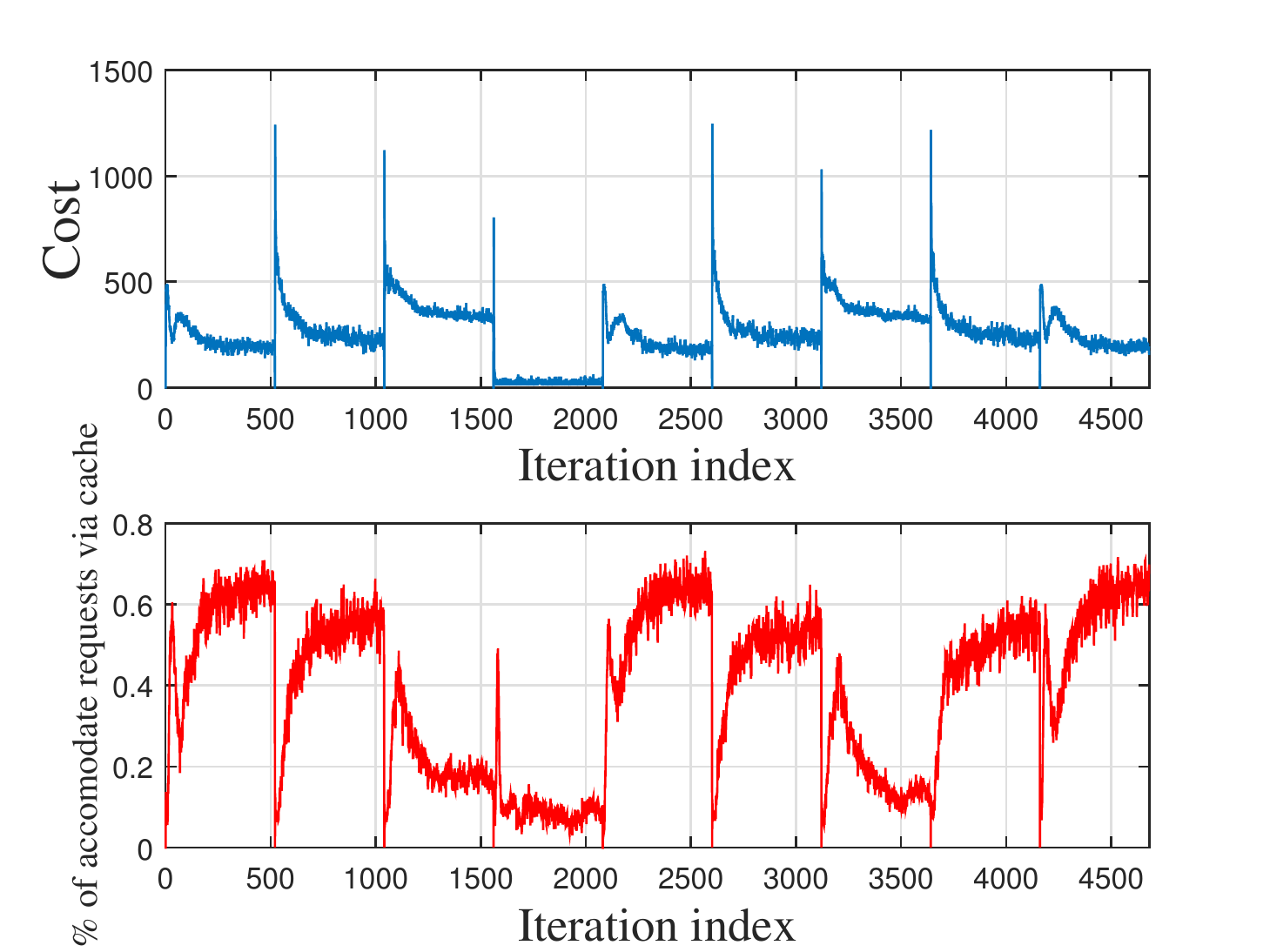}
		\caption{Caching performance.}
	\end{subfigure}
	\caption{Caching under dynamic cost parameters.}
	\label{dynamicprofile}
\end{figure}

{\noindent{\bf Remark 3.} In this section, it is assumed that the file  popularities demonstrate correlation across time, and thus do not change dramatically from one slot to the other. Thus, it is assumed that the popularity profile does not dramatically change within the interval of interest, and the realization of a large portion of possible states is considered extremely rare. Therefore, relatively small number of states, say 50 in our setting, is assumed to practically cover the most likely states.	 In a broader scenario, where a larger number of states are to be considered, continuous function approximation techniques such as kernel-based or deep learning approaches \cite{ormoneit2002kernel,lillicrap2015continuous} can be utilized to enable the algorithm with further scalability. }

{Finally, numerical tests  are carried to elaborate the impact of dynamic costs dictated to CCUs according to a cost parameter profile. The preselected profiles are reported in Fig.\ref{dynamicprofile}(a), and Fig.\ref{dynamicprofile}(b) shows corresponding cost and percentage of accommodated requests via cache. The two-state Markov chain for global and local popularity profiles are considered the same as described earlier in this section. As the percentage of accommodated requests via cache demonstrates, the caching policy in any interval is directly influenced by the corresponding cost parameters, and thus can be controlled via the network operator. }

\section{Conclusions}
The present work addresses caching in  5G cellular networks, where space-time popularity of requested files is modeled via local and global Markov chains. By considering local and global popularity mismatches as well as cache-refreshing costs, 5G caching is cast as a reinforcement-learning task. A Q-learning algorithm is developed for finding the optimal caching policy in an online fashion, and its linear approximation is provided to offer scalability over large networks. The novel RL-based caching offers an asynchronous and semi-distributed caching scheme, where adaptive tuning of parameters can readily bring about policy adjustments to space-time variability of file requests via light-weight updates.

\bibliographystyle{IEEEtran}
\bibliography{IEEEfull,biblio_pro_caching}

\end{document}